%% file: m-arxiv.tex
\documentclass{article}

\usepackage{arxiv}

\usepackage[utf8]{inputenc} % allow utf-8 input
\usepackage[T1]{fontenc}    % use 8-bit T1 fonts
\usepackage{hyperref}       % hyperlinks
\usepackage{url}            % simple URL typesetting
\usepackage{booktabs}       % professional-quality tables
\usepackage{amsfonts}       % blackboard math symbols
\usepackage{nicefrac}       % compact symbols for 1/2, etc.
\usepackage{microtype}      % microtypography
\usepackage{lipsum}
\usepackage{graphicx}
\usepackage{array}
\usepackage{caption}
\usepackage{cleveref}
\usepackage[dvipsnames]{xcolor}
\usepackage[numbers,sort&compress]{natbib}
\usepackage{caption}

\usepackage{listings}

\title{Chatbot Companionship: A Mixed-Methods Study of Companion Chatbot Usage Patterns and Their Relationship to Loneliness in Active Users}

\author{
 \textbf{Auren R. Liu} \\
  MIT Media Lab \& Harvard-MIT Health Sciences and Technology, \\
  Massachusetts Institute of Technology. \\
  Cambridge, MA 02139. \\
  \texttt{rliu34@media.mit.edu} \\[1em]
 \textbf{Pat Pataranutaporn} \\
  MIT Media Lab, \\
  Massachusetts Institute of Technology. \\
  Cambridge, MA 02139.  \\
  \texttt{patpat@mit.edu} \\[1em]
 \textbf{Pattie Maes} \\
  MIT Media Lab, \\
  Massachusetts Institute of Technology. \\
  Cambridge, MA 02139.
}

\begin{document}

\maketitle

\begin{abstract}
Companion chatbots offer a potential solution to the growing epidemic of loneliness, but their impact on users' psychosocial well-being remains poorly understood, raising critical ethical questions about their deployment and design. This study presents a large-scale survey (n = 404) of regular users of companion chatbots, investigating the relationship between chatbot usage and loneliness. We develop a model explaining approximately 50\% of variance in loneliness; while usage does not directly predict loneliness, we identify factors including neuroticism, social network size, and problematic use. Through cluster analysis and mixed-methods thematic analysis combining manual coding with automated theme extraction, we identify seven distinct user profiles demonstrating that companion chatbots can either enhance or potentially harm psychological well-being depending on user characteristics. Different usage patterns can lead to markedly different outcomes, with some users experiencing enhanced social confidence while others risk further isolation. These findings have significant implications for responsible AI development, suggesting that one-size-fits-all approaches to AI companionship may be ethically problematic. Our work contributes to the ongoing dialogue about the role of AI in social and emotional support, offering insights for developing more targeted and ethical approaches to AI companionship that complement rather than replace human connections.
\end{abstract}

\begin{figure}
  \includegraphics[width=\textwidth]{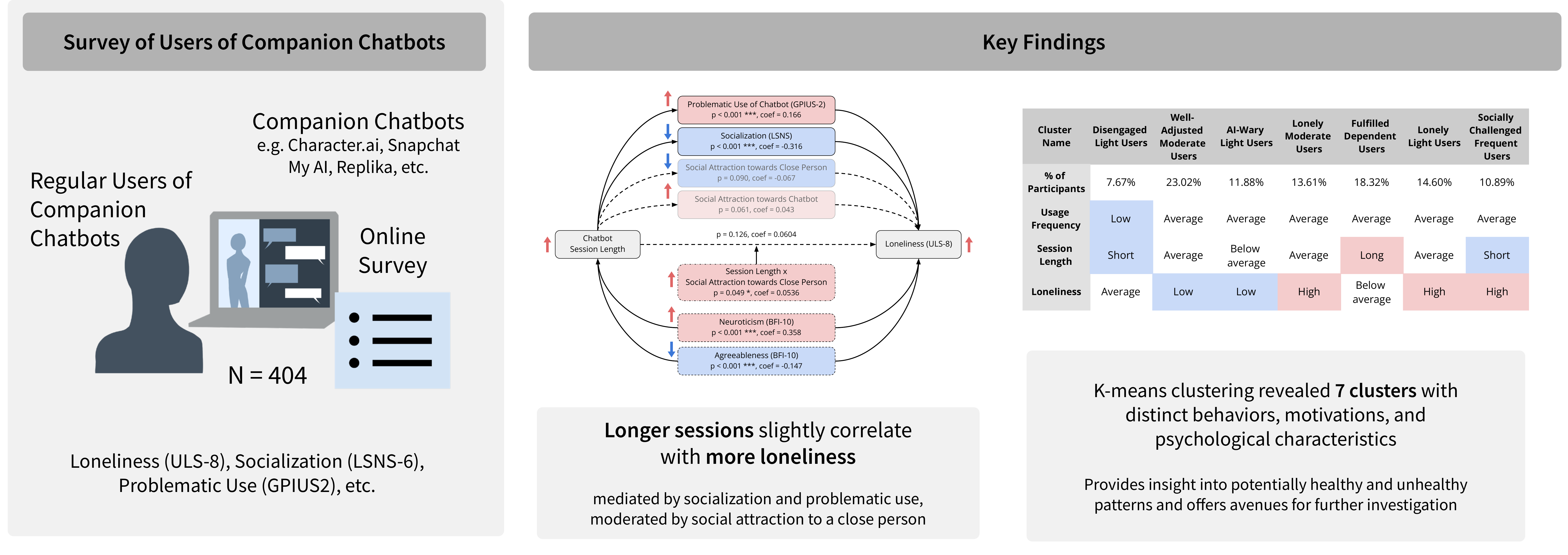}
  \caption*{\small \textbf{Overview} of the study’s methodology and major findings. A survey of 404 participants explored chatbot usage and its psychological impacts. From the findings, we developed a multiple regression model and identified seven user archetypes.}
  \label{fig:teaser}
\end{figure}

\section{Introduction}
In an increasingly digital world, we are paradoxically more connected and isolated than ever. Despite unprecedented technological means to connect, loneliness remains pervasive, affecting approximately \textbf{a third of individuals in industrialized countries} \cite{Loveys2019-oq}. This phenomenon has led to what the U.S. Surgeon General declared an ``epidemic of loneliness'' \cite{SG_loneliness-fb}. The gravity of this issue is underscored by research indicating that chronic loneliness can increase mortality risk by 26-29\%, comparable to smoking 15 cigarettes a day \cite{Chawla2021-ek}.  

Artificial intelligence (AI) has emerged as a potential tool for addressing loneliness. Companion chatbots, which engage in complex conversations and maintain persistent relationships with users, have gained significant popularity, with platforms like Replika and Character.ai reporting over 30 million and 20 million users, respectively \cite{Patel2024-wq, Shewale2024-sm}. These companion chatbots offer users the opportunity for constant, judgment-free interaction. Some research suggests that companion chatbots, even when used over the short term, decrease feelings of loneliness \cite{Loveys2019-oq, Gasteiger2021-bq, De-Freitas2024-hn}. However, concerns have been raised about the authenticity of these relationships, suggesting that they lower the bar for what people think relationships can be \cite{Turkle2011-dz}. This could lead to unrealistic expectations and further social withdrawal.

The nuanced nature of human-AI interaction in this context presents a critical gap in our understanding. Specifically, we lack comprehensive knowledge about factors influencing companion chatbots' impact on loneliness and other psychological and social factors. Artificial agents may complement human social interactions in some contexts while replacing them in others \cite{Xia2024-mv}. This variability in outcomes underscores the need for a more comprehensive and granular analysis of user characteristics, usage patterns, and their relationships to psychological well-being. To address this gap, we conducted a survey (n = 404) of regular companion chatbot users, with the following research questions: 

\begin{itemize}
    \item RQ1: What motivates the use of companion chatbots, and what do people use them for? 
    \item RQ2: What is the relationship between companion chatbot usage and loneliness, and what factors mediate or moderate this relationship?
    \item RQ3: Can we identify and characterize distinct user profiles based on usage patterns, loneliness levels, problematic use, and other psychological and social factors?  
\end{itemize}

Our study aims to provide a more nuanced understanding of the impact of companion chatbots. By identifying the characteristics of users who experience psychosocial benefits from chatbot interactions (i.e., reduced loneliness and improved social skills) and those who may experience negative consequences (i.e., emotional dependence and social withdrawal), we hope to inform more targeted and ethical design practices. For example, more information could guide the development of customizable interfaces that emphasize features beneficial to specific user groups or implement safeguards for at-risk users. By examining how users' mental models and expectations of companion chatbots influence outcomes, we can explore strategies for fostering healthy human-AI relationships that complement rather than replace human connections.

Through a combination of statistical analyses, including multiple regression and cluster analysis, along with qualitative assessment of user responses, we present the following contributions: (1) A \textbf{model} explaining approximately 50\% of variance in loneliness, identifying problematic use as a key mediator and social attraction as a moderator in the relationship between chatbot usage and loneliness while accounting for personality traits and social network characteristics; (2) a \textbf{typology} of seven distinct companion chatbot user profiles demonstrating how similar usage patterns can lead to markedly different outcomes based on individual characteristics and social contexts; (3) an exploration of \textbf{motivations and usage characteristics} across different user types, revealing how technological exploration and entertainment dominate over social purposes despite marketing as social tools; (4) a demonstration of how manual thematic analysis can be effectively triangulated with \textbf{LLM-based theme extraction} to achieve both interpretive depth and comprehensive coverage in large-scale qualitative analysis; and (5) \textbf{design implications} for creating more effective and ethically responsible companion chatbots, including the need for personalized approaches and built-in mechanisms to detect and intervene in potentially harmful usage patterns.

By addressing these questions, we aim to advance the field of human-computer interaction in the domain of AI companionship, providing insights that can guide the development of technologies that genuinely enhance human well-being and social connection in our increasingly digital world.

\section{Background and Related Work}

\subsection{Loneliness and Social Support}
Loneliness, the subjective feeling of isolation regardless of actual social network, affects approximately one-third of individuals in industrialized countries \cite{Loveys2019-oq}. The U.S. Surgeon General's 2023 declaration of loneliness as an epidemic highlights this issue's significance \cite{SG_loneliness-fb}. Chronic loneliness increases mortality risk by 26-29\%, comparable to smoking 15 cigarettes daily \cite{Freedman2020-qw}.

Individuals employ various coping strategies for loneliness \cite{Rokach1998-xg, Deckx2018-op}, with digital platforms emerging as common mechanisms. Evidence shows that lonely, socially isolated, and socially anxious individuals tend to engage in one-sided paradigms like parasocial relationships with content creators \cite{Niu2021-yr, de-Berail2019-og, Balcombe2023-kw, Madison2016-gy} and face a higher risk of addiction to online platforms \cite{Reer2019-hq, O-Day2021-mj, Burke2010-aw}.

Effective interventions for reducing loneliness include cognitive behavioral therapy, improving social support, and facilitating social interaction \cite{Rokach1998-xg, Gasteiger2021-bq, Hickin2021-xj}, though efficacy varies across individuals. This aligns with recognition that lonely individuals constitute a diverse group with varying needs \cite{Cacioppo2018-it}, suggesting interventions—including technological ones—must be tailored to individual circumstances.

\subsection{Companion Chatbots}
Companion chatbots represent a novel approach to addressing loneliness \cite{Wang2024-ca, Goodings2024-qa, Boine2023-xx, Chen2024-jp, Shani2022-xl, Brewer2022-tj}. These commercially available chatbots employ generative AI to engage in complex conversations and maintain persistent relationships with users. Several popular services have gained significant user bases: Replika reported over 30 million registered users worldwide by August 2024 \cite{Patel2024-wq}, while Character.ai, a platform allowing users to role-play with chatbots based on fictional characters, reported over 20 million users in March 2024 \cite{Shewale2024-sm}.

The popularity of these services, particularly among young adults, indicates their growing acceptance as sources of social interaction and emotional support \cite{Koulouri2022-ct, Xygkou2024-dt, Xygkou2023-ql}. Research suggests artificial companions can reduce loneliness through direct companionship, acting as catalysts for social interaction, facilitating remote communication, and providing reminders for social activities \cite{Gasteiger2021-bq}.

The perceived ``realness'' of connection with companion chatbots significantly impacts their effectiveness. Users who attribute human-like qualities to chatbots often experience improved social health and reduced loneliness \cite{Xia2024-mv}, though this carries a risk of dehumanizing other people \cite{Zehnder2021-vf}. Pataranutaporn \& Liu et al. \cite{Pataranutaporn2023-ca} found that mental models altered experience—priming users with different beliefs about an AI's motives significantly affected their assessment of the AI's trustworthiness and effectiveness.

Research on artificial agents' influence on human socialization has produced mixed results, mostly qualitative in nature \cite{Van_der_Loos2014-qh, Turkle2011-dz}. One study suggested artificial agents may replace humans for task-oriented interactions while complementing human social interactions \cite{Xia2024-mv}.

While empathy is inversely correlated with loneliness \cite{Beadle2012-lw}, the role of chatbots that perform empathy remains uncertain. Role-play increases empathy \cite{Bearman2015-py, Dewi2019-hn}, yet the potential of AI-facilitated role-play to influence empathy and loneliness represents an unexplored research area despite widespread user engagement.

\subsection{Human-AI Interaction for Emotional Support}
Studies suggest even limited AI companionship reduces loneliness \cite{De-Freitas2024-hn}, though long-term effects and risks remain subjects of investigation and ethical debate \cite{Li2023-ew, Abd-Alrazaq2020-ty}. Skjuve et al. described how human-chatbot relationships evolve from superficial interaction through affective exploration to stable emotional connection \cite{Skjuve2021-uf}. This enables companion chatbots to provide various forms of social support in an always-available ``safe space'' \cite{Ta2020-fu}. These interactions appeal particularly to those lacking traditional support systems, such as LGBTQ individuals \cite{Ma2024-ll}. User motivations range from curiosity to explicit seeking of emotional support, with topics spanning casual conversation to deep personal issues \cite{Ta-Johnson2022-xw}. Siemon et al.'s analysis of Replika reviews \cite{Siemon2022-lt} showed users often employ the service to cope with loneliness, with long-term interactions proving beneficial.

Significant ethical considerations include potential over-reliance on AI, consequences for mood, delays in seeking professional help, and withdrawal from human socialization \cite{Pataranutaporn2021-qc}. A teen suicide case in 2024 following extensive Character.ai use raised concerns about chatbot influence on mental health \cite{Roose2024-nw}, underscoring the need to understand companion chatbots' impact on psychosocial well-being.

Our research seeks to identify characteristics of users who benefit from chatbot interactions versus those who experience negative consequences, informing targeted and ethical design practices toward positive computing in human-AI interaction—designing technology that supports psychological well-being \cite{Calvo2017-br}.

\section{Methods}
We surveyed companion chatbot users recruited from CloudResearch Connect \cite{Hartman2023-hy}, administering psychological scales and measuring chatbot usage patterns. The study was IRB-approved with a target sample of 385 participants (95\% confidence level, 5\% margin of error). Participants were 18+ years old, English-fluent, and regular chatbot users (weekly use for at least one month). After filtering incomplete responses, our final sample consisted of 404 participants.

\subsection{Analytical Approach}
Our mixed-methods approach included (1) Spearman correlation analysis of numerical variables with Benjamini-Hochberg correction, (2) multiple regression exploring relationships between chatbot usage, loneliness, and potential mediating/moderating factors, (3) K-means clustering to identify distinct user profiles based on psychosocial characteristics and usage patterns, and (4) thematic analysis from a stratified random sample (N = 105) combined with generative AI-based topic extraction of the full dataset. 

\subsection{Survey Instrument}
Our survey combined established scales and custom items:

\subsubsection{Established Scales}
Our primary dependent variable was measured with the \textbf{UCLA Loneliness Scale} (ULS-8) \cite{Hays1987-jt}, selected for its brevity while maintaining strong psychometric properties (Cronbach's $\alpha$ = .84). For potential mediating and moderating factors, we included the following items: \textbf{Lubben Social Network Scale} (LSNS-6) \cite{Lubben2006-bw}, measuring social isolation ($\alpha$ = .83); \textbf{Multidimensional Scale of Perceived Social Support} (MSPSS) \cite{Zimet1990-go}, assessing perceived social support ($\alpha$ = .88); \textbf{Brief Rosenberg Self-Esteem Scale} (B-RSES) \cite{Monteiro2022-yi}, evaluating self-esteem ($\alpha$ = .89); \textbf{Big Five Inventory} (BFI-10) \cite{Rammstedt2007-uu}, assessing personality traits (mean correlation with full BFI: .83); \textbf{Attitudes Towards AI Scale} \cite{Schepman2020-us}, measuring AI attitudes (positive subscale: $\alpha$ = .88; negative: $\alpha$ = .83);
\textbf{Human-Chatbot Interaction Effect Scale} \cite{Xia2024-mv}, measuring effects of human-chatbot interaction on human-human interaction;
\textbf{Adapted Generalized Problematic Internet Use Scale} (GPIUS2) \cite{Caplan2010-yt}, measuring problematic chatbot use ($\alpha$ = .91 for original scale);
\textbf{State Empathy Scale} \cite{Shen2010-zb}, capturing empathy dimensions with strong construct validity;
\textbf{Interpersonal Attraction Scale} \cite{McCroskey1974-sz}, measuring attraction dimensions, with versions for chatbots and close persons;
\textbf{Perceived Homophily Scale} \cite{Mccroskey1975-bm}, measuring perceived similarity in communication;
\textbf{Attributional Confidence Scale} \cite{Gudykunst1986-hp}, assessing ability to predict others' attitudes and behaviors; and
\textbf{System Usability Scale} (SUS) \cite{Brooke1996-uk}, measuring perceived chatbot usability ($\alpha$ = .91).

Where necessary, we shortened scales based on factor loadings from previous studies to manage survey length while maintaining measurement integrity. We addressed this limitation through statistical controls for multicollinearity and conservative interpretation.

\subsubsection{Custom Items}
We developed custom items to collect data on chatbot usage frequency and session length, as well as usage motivations, conversation topics, and perceived impacts on relationships. Most items combined multiple-choice selections with free-response elaboration, with participants providing elaboration for reasons for \textbf{initially using} and \textbf{continuing to use} chatbots, conversation \textbf{topics}, perceived \textbf{effects of chatbot use on human relationships} (combined with Likert-scale items reporting changes in human interactions), situational \textbf{preferences for chatbot versus human} interaction (combined with a preference matrix across different contexts), and \textbf{reasons underlying these preferences}. 

\subsection{Statistical Analyses}
\begin{figure*}[t]
    \centering
    \includegraphics[width=1\linewidth]{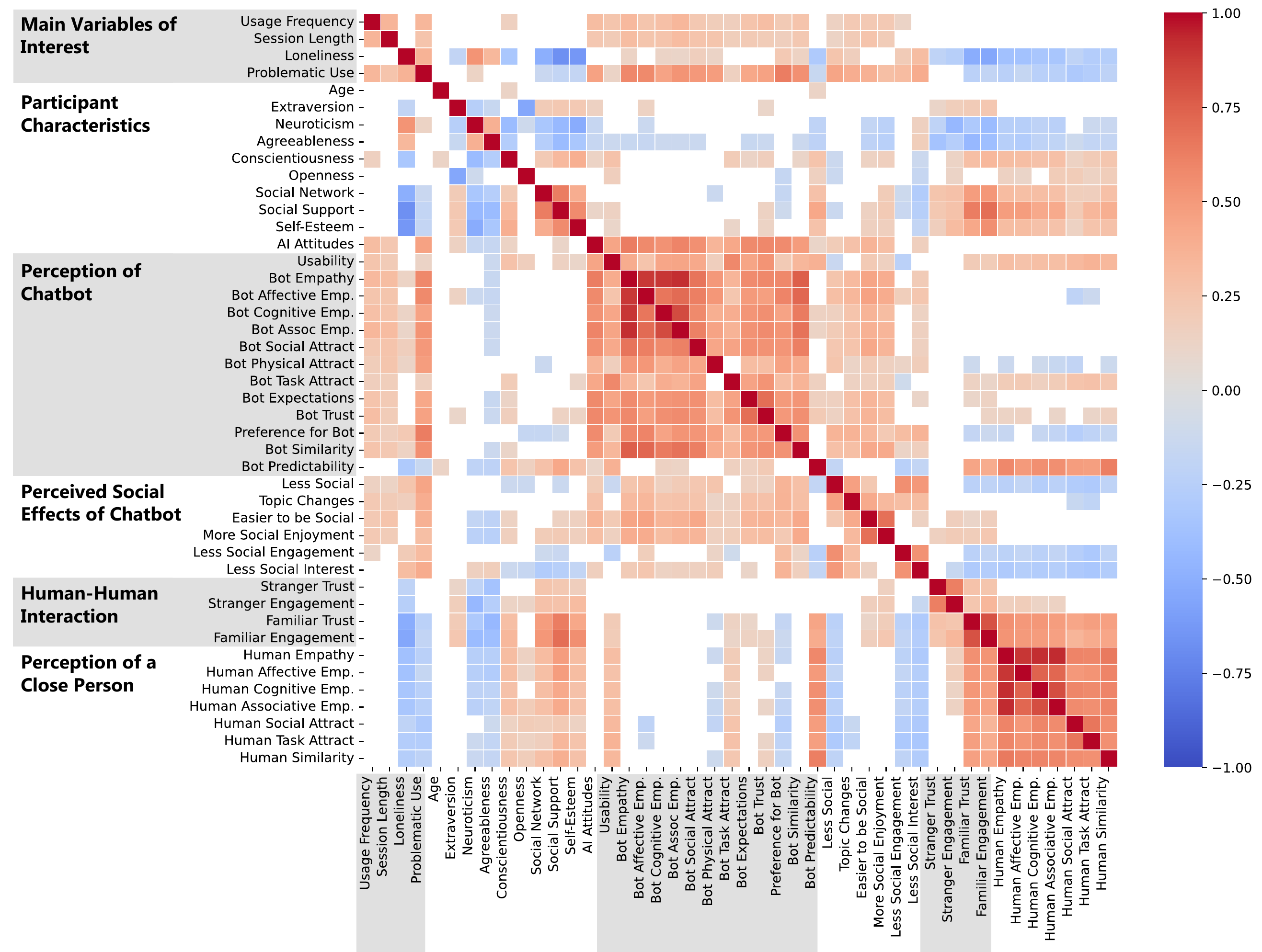}
    \caption{Correlation matrix (Spearman, Benjamini-Hochberg corrected) showing significant ($p < 0.05$) correlations of psychological, social, and usage characteristics. Variables are sorted into general categories, and names have been altered for readability. The original names can be seen in \Cref{fig:corrMatrixFull} of the Appendix.}
    \label{fig:corrMatrix}
\end{figure*}

All analyses were performed using Python. We calculated descriptive statistics on chatbot usage characteristics for both the entire dataset and identified clusters. Usage frequency and session length were converted to numerical scores, and all variables were standardized. For correlation analysis, we computed a Spearman correlation matrix with Benjamini-Hochberg correction (\Cref{fig:corrMatrix}). 

Regression analysis focused on \textbf{session length} as the independent variable (as frequency showed no significant correlation with loneliness). Potential mediators were identified using corrected p-values ($p < 0.05$) following Baron and Kenny's approach \cite{Baron1986-dj} with Sobel tests \cite{Sobel1982-hs}. Moderation effects were tested by examining interaction terms \cite{Aiken1991-am}. Variance Inflation Factors were calculated to address multicollinearity \cite{Hair2010-ir, O-brien2007-nk}. Our regression model initially included all identified mediators, moderators, and confounders \cite{Greenland1999-rx, VanderWeele2013-pn}. We used backward elimination to create a more parsimonious model \cite{Cohen2003-qt}, retaining age and sex as control variables. Model fit was assessed using R-squared, F-statistic, and condition number, with assumptions checked using Q-Q plots, residuals vs. fitted plots, and influence plots \cite{Cook1982-js, Fox2015-um}. We conducted parallel multiple mediation analysis using bootstrapping with 5,000 resamples \cite{Preacher2008-pa}.

We then employed \textbf{K-means clustering} to identify distinct user profiles. Features were selected based on significant correlations with key variables ($p > 0.05$, corr $> 0.3$) and relevance to chatbot usage, then reduced to 10 principal components. The optimal number of clusters was determined using the elbow method \cite{Kodinariya2013-rc}, silhouette scores \cite{Kaufman2009-ll}, and interpretability considerations. We chose a 7-cluster solution based on these metrics and the distinctiveness of resulting profiles. To validate our clustering, we conducted Kruskal-Wallis tests \cite{Kruskal1952-sl} followed by Dunn's post-hoc tests with Bonferroni correction \cite{Dunn1964-df}, calculated average silhouette scores \cite{Rousseeuw1987-pq}, and performed MANOVA \cite{Huberty2007-ua} with Box's M-test \cite{Box1949-ry}.

\subsection{Mixed-Methods Thematic Analysis}
To complement quantitative findings, we employed a mixed-methods approach combining manual thematic analysis \cite{Braun2006-qm} with automated topic extraction, addressing limitations of purely manual analysis (limited sample coverage) and purely automated analysis (reduced contextual sensitivity) \cite{O-Connor2020-gi}. Thematic analysis was conducted on a stratified random sample (N=105, 15 per cluster), with one researcher developing codes and grouping them into broader themes. To validate themes across the complete dataset (N=404), we implemented an \textbf{automated pipeline using GPT-4 for topic extraction} (see Appendix for details), processing responses in batches of 300 with instructions to identify 5-10 key themes per batch, including descriptions, prevalence estimates, and representative quotes. A researcher reviewed the LLM output to assess accuracy and distinctiveness before \textbf{comparing to manual themes} based on descriptions and quotes; LLM themes were matched to similar manual themes or identified as subsets where appropriate, with substantial overlap in identified patterns. This triangulation strengthened confidence in our thematic findings while demonstrating how manual depth can complement automated breadth in qualitative analysis.

\section{Results}

\subsection{RQ1: What motivates the use of companion chatbots, and what do people use them for?}

\begin{figure*}[t]
    \centering
    \includegraphics[width=0.8\linewidth]{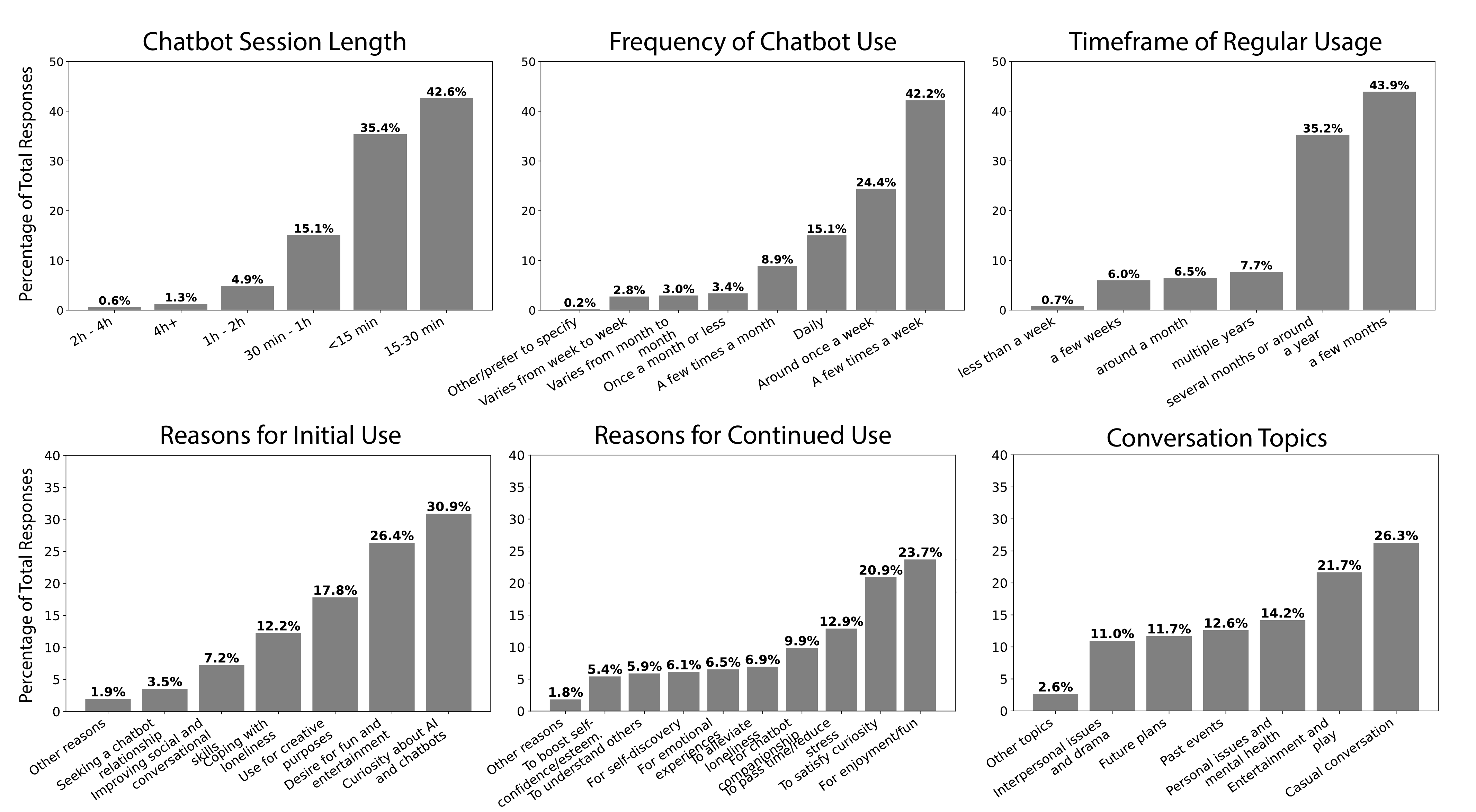}
    \caption{Percentage of total responses for multiple-choice questions regarding chatbot usage patterns among participants.} 
    \label{fig:chatbot-stat-1}
\end{figure*}

\subsubsection{Motivations for initial interest and continued use of companion chatbots}

Participants were asked separate questions about their initial interest in starting chatbot use and their reasons for continuing use, each combining multiple-choice options with free-response elaboration. Analysis of quantitative and qualitative data revealed diverse motivations, validated through automated theme extraction across the full dataset. Descriptions of these themes can be found in \Cref{fig:cluster_themes} in the Appendix. 

\textbf{Technology exploration} emerged as the predominant motivation (30.89\% initial use, 20.91\% continued use), with users expressing curiosity about AI capabilities: ``I wanted to see if it could actually emulate a real person.''

\textbf{Recreation} represented another major theme, encompassing entertainment (26.35\% initial, 23.69\% continued) and creative activities (17.81\% initial use). Users enjoyed role-playing and creative projects: ``I use chatbots to practice conversations for my D\&D game, creating NPC personalities and dialogue.''

\textbf{Practical utility} emerged particularly in continued use, with users valuing task assistance and information seeking: ``I usually use chatbots to help with ideas like dinner ideas or information I may want to find out without using a typical search engine.''

The \textbf{safe space} theme encompassed chatbots as non-judgmental environments for self-expression: ``I sometime feel lonely and just want to be left alone, during this time I like chatting with my AI companion because I feel safe and won't not [sic] be judged.''

Notably, while companion chatbots are often marketed as solutions for loneliness, only 12.24\% of participants initially sought them for companionship (9.86\% continued). Other social motivations were also less prevalent but significant for some users, including \textbf{self-improvement} for social skills (7.24\% initial) and \textbf{lifestyle integration} where chatbots become part of daily routines. While initial interest often stems from curiosity and entertainment, continued use appears more rooted in practical and emotional benefits. 

\subsubsection{Topics of conversation with companion chatbots}
Analysis revealed five major themes in chatbot engagement  (see \Cref{fig:cluster_themes} in the Appendix). Our automated analysis confirmed these themes while providing prevalence estimates across all 404 participants. The convergence between manual and automated approaches strengthened confidence in our thematic findings.

\textbf{Casual exchange}, including casual conversations (26.28\%) and entertainment (21.66\%), represents the most common interaction: ``I often chat with my AI about random things like my day, and it feels like I'm having a normal conversation without the pressure.'' 

\textbf{Emotional disclosure} also emerged as notable, encompassing mental health discussions (14.17\%) and interpersonal issues (10.96\%): ``I can talk about my problems, and it's like having a private conversation with no fear of being criticized.''

\textbf{Knowledge seeking} involved users turning to chatbots for information and advice, including discussion of future plans (11.70\%). One individual mentioned, ``I use the chatbot to ask about random things like historical facts or recommendations for books.'' Others used companion chatbots as tools for specific tasks, such as recipe ideas. 

The theme of \textbf{creative development} manifested through topics of writing, roleplay, and creative ideation, with users leveraging chatbots as collaborators: ``I typically ask for ideas for my marketing art! I do a lot of graphic design for a lot of different people and sometime run low on creative ideas.'' A rare theme was \textbf{intimate exchange}, involving romantic and sexual interactions. 

This diverse range of conversation topics highlights the versatility of companion chatbots in meeting users' needs for entertainment, emotional support, and intellectual stimulation. 

\begin{figure*}
    \centering
    \includegraphics[width=1\linewidth]{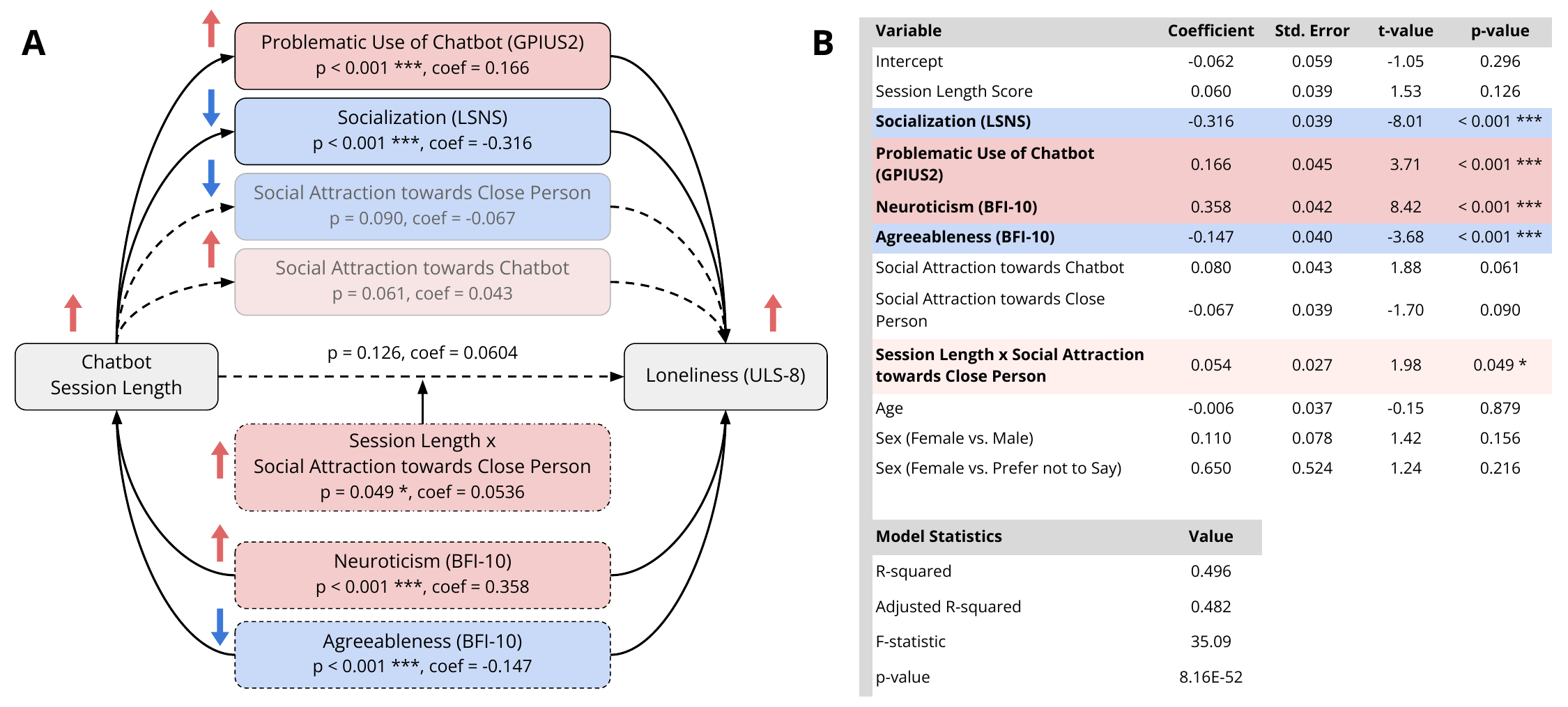}
    \caption{\textbf{A:} The final model from our exploratory multiple regression, depicting the relationships between chatbot session length, loneliness, and related factors. \textbf{B:} Model summary of an ordinary least squares regression for the effect of chatbot session length on loneliness (ULS-8), using standardized variables.}
    \label{fig:reg-model}
\end{figure*}

\subsection{RQ2: What is the relationship between companion chatbot usage and loneliness, and what factors mediate or moderate this relationship?}

\begin{figure}
    \centering
    \includegraphics[width=0.9\linewidth]{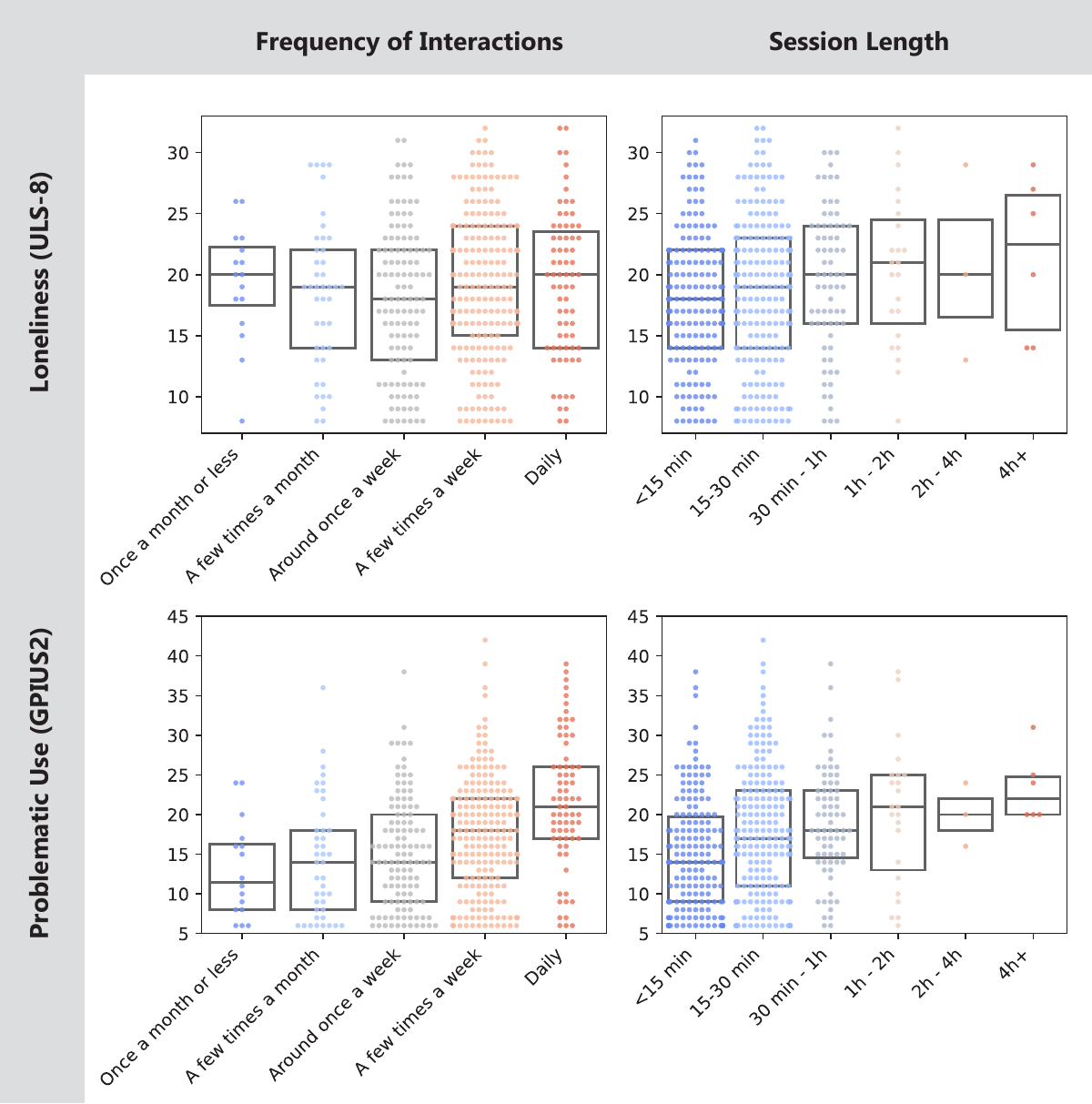}
    \caption{Plots of loneliness (ULS-8) and problematic usage (GPIUS2) vs. usage (frequency and session length). Top left: loneliness vs. chatbot usage frequency. Top right: loneliness vs. chatbot session length. Bottom left: problematic use vs. chatbot usage frequency. Bottom right: problematic use vs. chatbot session length.}
    \label{fig:chatbot-length}
\end{figure}

\subsubsection{Correlations in psychological, social, and companion chatbot usage characteristics}
For our prespecified analyses of usage vs. loneliness, we found a small but significant correlation between chatbot session length and loneliness ($r = 0.101, p = 0.042$), but not between loneliness and usage frequency ($r = 0.053, p = 0.287$), as shown in \Cref{fig:chatbot-length}.

Several other notable correlations emerged in our exploratory analysis using Benjamini-Hochberg corrected p-values. Chatbot usage (both length and frequency) showed moderate positive correlations with state empathy towards chatbots ($r > 0.3, p < 0.001$) and various forms of attraction towards chatbots ($r > 0.18, p < 0.001$). Loneliness demonstrated significant correlations with personality traits of extraversion, ($r = -0.188, p < 0.001$), conscientiousness ($r = -0.335, p < 0.001$), agreeableness ($r = -0.333, p < 0.001$), and neuroticism ($r = 0.526, p < 0.001$), aligning with Buecker et al.'s \cite{Buecker2020-bw} meta-analysis. Additionally, problematic use of chatbots (GPIUS2) showed moderate correlations with both session length ($r = 0.303, p < 0.001$) and frequency ($r = 0.337, p < 0.001$). See \Cref{fig:corrMatrix} for a correlation matrix of our continuous variables.

While these correlations provide initial insights into the relationships between variables, they represent associations rather than causation. A more detailed examination of these relationships through multiple regression analysis follows.

\subsubsection{Multiple Regression Model} 
We conducted an ordinary least squares (OLS) regression analysis to examine the relationship between session length and loneliness while controlling for other factors. The model explained a significant proportion of the variance in loneliness (\(R^2 = 0.496\), adjusted \(R^2 = 0.482\), \(F(11, 392) = 35.09\), \(p < .001\)) (\Cref{fig:reg-model} and \Cref{fig:reg-model}).

While session length alone did not significantly predict loneliness (\(\beta = 0.060\), \(p = 0.126\)), neuroticism (\(\beta = 0.358\), \(p < .001\)) and problematic chatbot use (\(\beta = 0.166\), \(p < .001\)) showed positive associations with loneliness. Social network size (\(\beta = -0.316\), \(p < .001\)) and agreeableness (\(\beta = -0.147\), \(p < .001\)) showed negative associations. We found a significant interaction effect (see \Cref{fig:interactions} in the Appendix) between session length and social attraction to a close person (\(\beta = 0.054\), \(p = 0.049\)). This interaction suggests that for individuals with higher social attraction to close others, longer chatbot sessions were associated with slightly increased loneliness, while this relationship was absent or reversed for those with lower social attraction to close others.

A parallel multiple mediation analysis (\Cref{fig:mediation} in the Appendix) revealed that problematic internet use (GPIUS2) significantly mediated the relationship between session length and loneliness ($\beta = 0.0183$, 95\% CI $[0.0027, 0.0403]$), while social network size (LSNS) did not show significant mediation ($\beta = 0.0007$, 95\% CI $[-0.0304, 0.0340]$). The total indirect effect was not statistically significant ($\beta = 0.0191$, 95\% CI $[-0.0149, 0.0563]$).

% ULS_8_centered ~ Session_Length_Score_centered + C_Social_Attract_centered + GPIUS2_centered + LSNS_Score_centered + BFI_Agreeableness_centered + H_Social_Attract_centered + BFI_Neurotic_centered + Sex + Age_centered + SLS_x_H_Social_Attract_centered

\subsection{RQ3: Can we identify and characterize distinct user profiles based on usage patterns, loneliness levels, problematic use, and other psychological and social factors?}

\subsubsection{Overall characteristics}

Participants ranged in age from 18 to 73, primarily 30-35. Most (93.6\%) were from the United States, white/Caucasian (66.3\%), and employed full-time (61.6\%), with a broad distribution of household income. For relationship status, most were single (42.1\%) or married (32.4\%). 59.7\% of participants identified as men and 36.6\% identified as women. Additional demographic characteristics are in \Cref{fig:demographs}.

Most participants utilize Snapchat MyAI (31.6\%), Character.ai (30.4\%), or Replika (22.6\%) and tend to spend 15 and 30 minutes in each session with a companion chatbot (42.6\%), with 35.4\% reported sessions shorter than 15 minutes. Very few users spend extended periods chatting, with only 1.28\% having sessions exceeding 4 hours. Most participants have been using their companion chatbots for a few months (43.9\%) to around a year (35.2\%), with only 7.7\% of participants having used their chatbots for multiple years and 13.2\% for around a month or less. More details on usage can be seen in \Cref{fig:chatbot-stat-1}. 

\subsubsection{Clusters of users of companion chatbots}
\begin{figure*}[t]
    \centering
    \includegraphics[width=0.8\linewidth]{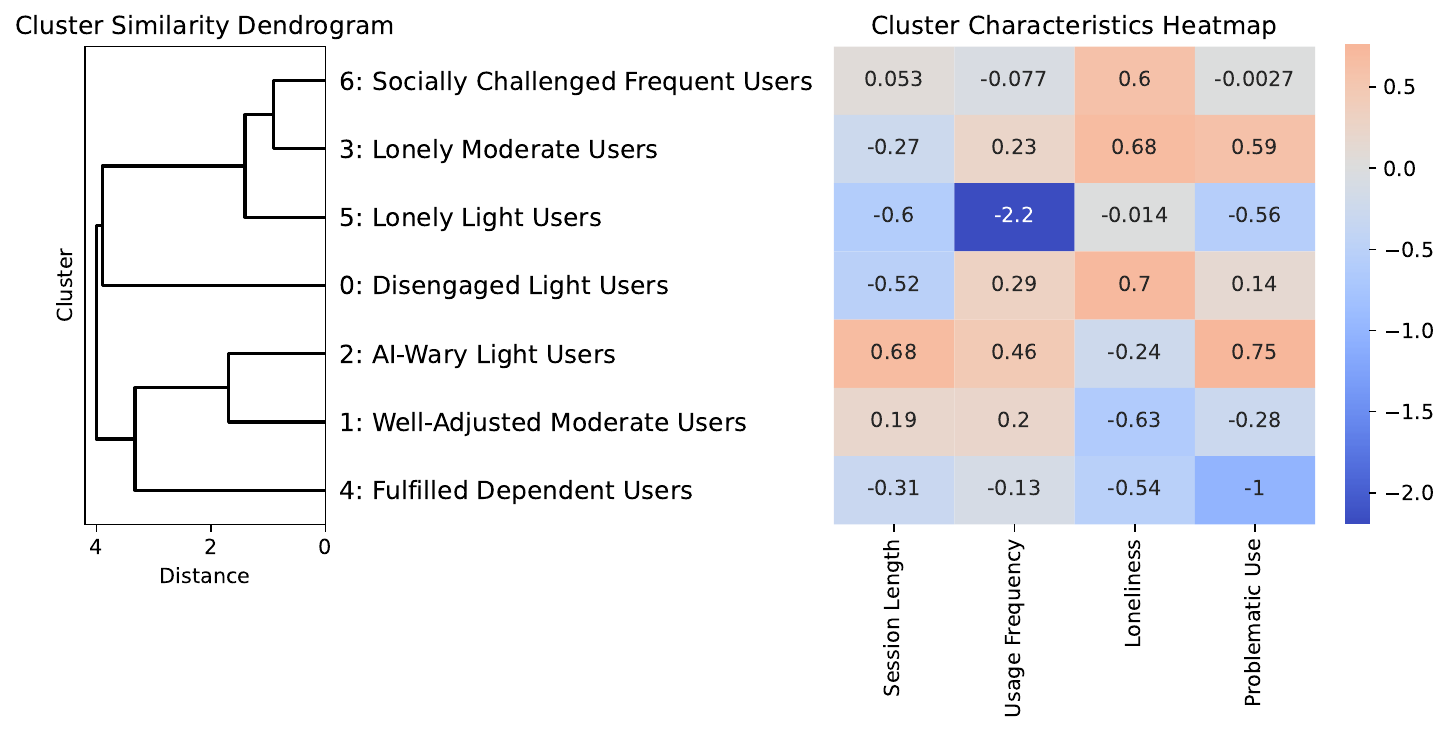}
    \caption{\textbf{Left:} a dendrogram depicting the degree of similarity between each cluster. \textbf{Right:} Cluster means of standardized variables of chatbot usage (frequency and session length), loneliness, and problematic use of chatbots for our seven user clusters. The cluster order corresponds to that of the dendrogram.}
    
    \label{fig:clusters}
\end{figure*}

Our K-means cluster analysis revealed seven distinct user profiles among companion chatbot users, each exhibiting unique behaviors, motivations, and psychological characteristics. The resulting clusters are visualized in the heatmap and dendrogram presented in \Cref{fig:clusters}. The demographics of each cluster, compared to the overall demographics, are shown in \Cref{fig:demographs}, with percentages provided in \Cref{fig:demographs} of the Appendix.

Validation of the cluster solution using Kruskal-Wallis tests revealed significant differences among clusters in terms of session length, loneliness (ULS-8), problematic use of chatbots (GPIUS2), and frequency of use (all $p < 0.001$). The average silhouette score ($-0.0413$) suggested some overlap between clusters, which is not uncommon in complex psychological data. A MANOVA test confirmed significant multivariate differences among the clusters when considering all variables simultaneously (Wilks' $\lambda = 0.2420, F(24, 1375.71) = 28.77, p < 0.001$). All four MANOVA test statistics (Wilks' lambda, Pillai's trace, Hotelling-Lawley trace, and Roy's greatest root) were significant ($p < 0.001$), providing robust evidence for cluster differences across multiple dimensions. Box's M-test indicated significant heterogeneity of covariance matrices ($\chi^2 = 181.96$, $p < 0.001$), which does not invalidate the clustering but suggests the need for cautious interpretation. This heterogeneity may affect the reliability of post-hoc comparisons and suggests that the relationship between variables varies across clusters. While these results provide strong support for the distinctiveness of the identified clusters, they should still be considered exploratory, providing a foundation for more targeted investigations in future research.

The seven identified clusters represent distinct patterns of chatbot engagement and psychological characteristics:

\begin{itemize}
\item \textbf{0: Disengaged Light Users} (7.67\%): Minimal chatbot engagement with average loneliness levels. Show neutral attitudes toward AI technology and average social support levels. Primary usage is brief (less than 15 min) casual conversations, with main motivations being curiosity to start use and practical utility for continued use. Average age is 32; for gender, 65\% are men and 35\% are women. 

\item \textbf{1: Well-Adjusted Moderate Users} (23.02\%): Average usage with low loneliness. Demonstrate high extraversion, low neuroticism, and strong social networks. Show positive attitudes toward both AI and human interactions. Main motivations of curiosity to start use and emotional support to continue use. Average age is 36; 67\% men, 33\% women.

\item \textbf{2: AI-Wary Light Users} (11.88\%): Average to low usage with low loneliness. Show high skepticism toward AI technology while maintaining strong human connections. Users were interested in chatbots for practical utility and recreation, with continued use motivated by practical purposes, such as seeking knowledge and advice. Average age is 34; 65\% men, 35\% women.

\item \textbf{3: Lonely Moderate Users} (13.61\%): Average usage with high loneliness. Show low socialization but positive attitudes toward chatbots. Notable for seeking emotional support and companionship through chatbot interactions, with some participants integrating chatbots into their lifestyle. Average age is 35; 51\% men, 45\% women, 4\% other.

\item \textbf{4: Fulfilled Dependent Users} (18.32\%): High usage with below-average loneliness. Show strong emotional engagement with chatbots while maintaining adequate social connections. Report positive impacts on real-world social interactions. Average age is 35; 61\% men, 37\% women, 1\% other.

\item \textbf{5: Lonely Light Users} (14.60\%): Average usage with high loneliness. Demonstrate high neuroticism and low self-esteem. Users seek out chatbots for emotional support and companionship, often discussing personal issues and sometimes engaging in intimate relationships with chatbots. Average age is 34; 47\% men, 52\% women.

\item \textbf{6: Socially Challenged Frequent Users} (10.89\%): Average frequency but short sessions, with high loneliness. Show low extraversion and social support, with notably low trust in both familiar and unfamiliar people. Both initial and continued use is motivated by a desire for companionship, with a few rare cases engaging in intimate relationships with their chatbot. Casual exchange was a less common theme of conversation. Average age is 36; 61\% men, 39\% women.
\end{itemize}

Detailed heatmaps including all numerical scales can be seen in \Cref{fig:detailedclusterheatmap}. A version with the original values of the scales can be found in \Cref{fig:og_detailedclusterheatmap} along with detailed cluster descriptions in the Appendix. The patterns of themes for each cluster's motivations for use and topics of discussion can be seen in \Cref{fig:cluster_themes} in the Appendix.

Post-hoc analysis using Dunn's test with Bonferroni correction revealed distinct patterns among clusters across our key variables (\Cref{fig:cluster_comparisons}). For \textbf{session length}, Fulfilled Dependent Users (Cluster 4) showed significantly longer sessions than all other clusters ($p < .001$) except Well-Adjusted Moderate Users. Disengaged Light Users (Cluster 0) had significantly shorter sessions than most other clusters ($p < .001$).

\textbf{Usage frequency} showed less variation between clusters, with only Disengaged Light Users showing significantly lower frequency than all other clusters ($p < .001$), and Fulfilled Dependent Users showing higher frequency than AI-Wary Light Users and Lonely Light Users ($p < .01$).

For \textbf{loneliness} (ULS-8), three clusters - Lonely Moderate Users, Lonely Light Users, and Socially Challenged Frequent Users (Clusters 3, 5, and 6) - showed significantly higher levels than Well-Adjusted Moderate Users and AI-Wary Light Users ($p < .001$). Fulfilled Dependent Users maintained significantly lower loneliness levels than these high-loneliness clusters ($p < .001$).

\textbf{Problematic use} (GPIUS2) was highest in Lonely Moderate Users and Fulfilled Dependent Users (Clusters 3 and 4), who showed significantly higher levels than all other clusters ($p < .001$) except each other. AI-Wary Light Users (Cluster 2) demonstrated the lowest levels of problematic use ($p < .001$).  

\section{Discussion}

Our study examines the relationship between companion chatbot usage and loneliness, revealing complex interactions among psychological, social, and technological factors.

\subsection{Understanding the Relationship Between Companion Chatbot Usage and Loneliness}

For RQ2, our regression analysis showed no significant direct relationship between session length and loneliness, but revealed important mediating and moderating factors. Problematic use emerged as a significant mediator, suggesting increased usage creates more opportunities for problematic behaviors, aligning with research connecting problematic internet use with loneliness \cite{Kim2009-ef}. 

Usage is negatively associated with social network size, though causality remains unclear—users might reduce human interactions as they spend more time with chatbots or turn to chatbots to fill existing social voids. The positive relationship between usage and social attraction towards chatbots is logically consistent, as interaction time allows deeper attachments to form.

Among confounding variables, neuroticism showed strong positive association with loneliness, while agreeableness demonstrated strong negative association, aligning with previous research \cite{Buecker2020-bw} and reinforcing the importance of controlling for these factors.

The positive interaction between session length and social attraction towards close people suggests a nuanced relationship, possibly indicating a mismatch between social needs and chosen interactions, where time with chatbots may be perceived as time not spent with valued human connections.

\subsection{Understanding Clusters of Chatbot Users}

For RQ1, most participants used companion chatbots primarily for technological exploration and entertainment rather than friendship or intimate relationships. Our cluster analysis for RQ3 revealed distinct user profiles demonstrating complex relationships with these technologies.

The most striking contrast emerged between \textbf{Socially Fulfilled Dependent Users} (Cluster 4) and \textbf{Lonely Moderate Users} (Cluster 3). Both exhibit high problematic use, yet with dramatically different outcomes. Socially Fulfilled Dependent Users maintain high well-being despite extensive chatbot engagement, paralleling findings that intensively involved gamers experience psychosocial benefits from their gaming \cite{Snodgrass2018-zb}. This challenges concerns about AI replacing human relationships \cite{Pataranutaporn2021-qc, Turkle2011-dz}. Conversely, Lonely Moderate Users with similar usage patterns experience high loneliness, low socialization, and preference for chatbots over humans.

\textbf{Socially Challenged Frequent Users} (Cluster 6) and \textbf{Lonely Light Users} (Cluster 5) present another interesting comparison. Both seek chatbots for companionship, with participants discussing romantic or sexual relationships with their chatbots. These clusters most strongly reflect the Media Equation \cite{Reeves1998-nr}. Of concern is whether Lonely Light Users might transition to Socially Challenged Frequent Users over time, losing trust and empathy with people. Understanding how these users' characteristics change over time may be valuable for future research.

\textbf{Well-Adjusted Moderate Users} (Cluster 1), the largest group, demonstrate a potentially optimal engagement pattern, maintaining strong human connections while using chatbots primarily as tools for emotional processing. Many disclose emotional issues to chatbots, valuing the safe space they provide. Recent work showing social support as a significant mediator between psychological problems and problematic gaming \cite{Malak2023-nt} may help explain how Cluster 3, with poor perceived social support, exhibits high loneliness compared to Cluster 1.

The \textbf{AI-Wary Light Users} (Cluster 2) provide an important counterpoint, showing that emotional distance from chatbots can coexist with low loneliness. These users show low emotional attachment to chatbots and use them mostly for practical purposes, suggesting emotional investment in companion chatbots may not be necessary or optimal for all users.

\subsection{Ethical Considerations \& Design Implications}
Our findings raise critical questions about responsible companion chatbot development. The identification of vulnerable user populations who may develop problematic usage patterns highlights the need for proactive ethical safeguards. The contrast between users who benefit from AI companionship and those who may be harmed underscores that one-size-fits-all approaches may be ethically insufficient.

\textbf{Protecting Vulnerable Users.} Clusters 3, 5, and 6 represent particularly vulnerable populations who often turn to AI companions for emotional support and companionship, yet may be most susceptible to negative outcomes. Current platforms typically lack mechanisms to identify such users or provide appropriate interventions. Our findings suggest the need for \textbf{risk assessment tools} that identify users showing patterns associated with problematic use, \textbf{adaptive interfaces} that promote healthy usage patterns for vulnerable users, and \textbf{intervention mechanisms} such as usage limits, mental health resource suggestions, or encouragement of human social contact.

\textbf{Developer Responsibility \& Regulation.} Unlike traditional software tools, AI companions can explicitly encourage emotional attachment and dependency, creating ethical obligations similar to those in healthcare or counseling contexts. This includes implementing safeguards for user design principles focused on encouraging and facilitating human social connections. The largely unregulated nature of AI companion platforms, combined with their potential for both benefit and harm, suggests the need for evidence-based policy consideration. Our findings could inform regulatory approaches that balance innovation with user protection, particularly for vulnerable populations.

\subsubsection{Theoretical Contributions}

Our findings extend beyond simple social displacement theories \cite{Kraut1998-gl}, demonstrating that companion chatbots serve \textbf{different functions based on user characteristics and usage patterns}. This aligns with the Parasocial Contact Hypothesis \cite{Schiappa2005-us}, which suggests media characters can provide meaningful social experiences. The mediating role of problematic use suggests a pathway where increased chatbot engagement creates more opportunities for unhealthy usage patterns, which in turn exacerbate loneliness. This aligns with compensatory internet use theory \cite{Kardefelt-Winther2014-eb}, where individuals with social deficits may turn to technology to fulfill unmet social needs. However, when this compensatory use becomes problematic—loss of control, negative mood when unable to access—it may paradoxically worsen the loneliness it was meant to address. Extensive chatbot use may not be harmful if problematic patterns are avoided, as evidenced by our Fulfilled Dependent Users who maintain high engagement without increased loneliness. For those with strong existing social networks, companion chatbots may serve as \textbf{beneficial supplements} to human social interaction. However, socially isolated individuals risk problematic use \cite{Reer2019-hq, O-Day2021-mj}, and we must consider the risk of chatbots replacing human interactions and potentially exacerbating loneliness. The key finding is that similar usage patterns can lead to markedly different outcomes depending on user characteristics. We also demonstrate how \textbf{manual thematic analysis can be effectively combined with LLM-based theme extraction} to achieve both interpretive depth and comprehensive coverage in large-scale qualitative research, offering a scalable approach for mixed-methods studies.

\subsection{Limitations and Future Research}

The self-reported measures of our survey risk recall bias. Our custom instruments offered broad options to reduce inaccuracies but resulted in coarse measures. Our adapted scales were based on factor loadings from previous studies but may affect validated psychometric properties. To reduce participant fatigue, we prioritized shorter scales over maintaining full lengths. Our participants were primarily white/Caucasian Americans, limiting generalizability. Additionally, while statistically grounded, \textbf{cluster numbers} involve subjective judgments of interpretability. The heterogeneity of covariance matrices indicated by Box's M-test requires caution in interpretation. Our \textbf{exploratory regression model selection} warrants cautious interpretations as well. Assumption of linear relationships may not capture complex nonlinear relationships. The numerous predictors increase the risk of Type II errors \cite{Lavery2019-nt}. Our cross-sectional design limits causal inferences.

Our triangulated approach combining manual and automated thematic analysis strengthened theme validation, but \textbf{LLM-based analysis may miss subtle contextual nuances and interpretive depth} that human analysts capture. The automated approach also relies on the quality and consistency of the underlying language model, which may introduce systematic biases in theme identification or prevalence estimation. Another limitation is that our study identifies vulnerable populations, but we did not assess whether participants were currently \textbf{receiving mental health treatment} or had been advised against using AI companions, which could affect the interpretation of our findings regarding vulnerable users.

Several key areas warrant investigation: longitudinal studies to understand causal relationships; experimental studies manipulating specific chatbot design aspects; investigation of non-linear relationships; cross-cultural studies; and research on long-term effects on social skills and relationship expectations.

\section{Conclusion}

This study examines the relationship between companion chatbot usage and loneliness, revealing complex interactions among psychological, social, and technological factors. Our model explaining approximately 50\% of variance in loneliness demonstrates that while chatbot session length does not directly predict loneliness, this relationship is mediated by problematic use and moderated by social attraction to close others, with personality traits and social network characteristics serving as key confounding variables.

Our cluster analysis identified seven distinct user profiles demonstrating how similar usage patterns can lead to markedly different outcomes. Fulfilled Dependent Users maintain healthy relationships despite intensive chatbot use, while Lonely Moderate Users with similar patterns show signs of social withdrawal, emphasizing the importance of individual characteristics and social contexts in determining outcomes. Most participants used companion chatbots primarily for technological exploration and entertainment rather than social purposes, suggesting these tools serve broader functions than typically assumed. However, distinct vulnerable populations emerged who rely on chatbots for companionship and emotional support, raising important ethical considerations.

The ethical implications of our findings extend beyond individual user experiences to broader questions about technology's role in human social development. The identification of distinct user profiles and key mediating factors suggests the need for personalized approaches and built-in mechanisms to detect potentially harmful usage patterns. As AI companions become increasingly sophisticated and widespread, the need for responsible development practices becomes more urgent. Our research provides an empirical foundation for moving beyond abstract ethical debates toward concrete, evidence-based approaches to AI companion design that prioritize user well-being and social health.

\section{Acknowledgments}

Statistical support was provided by data science specialist Jinjie Liu, at IQSS, Harvard University.

\bigskip

\bibliographystyle{unsrt}
\bibliography{aaai25}

\appendix

\include{m_appendix}

\end{document}

%% file: m_appendix.tex
\section{Appendix}

\section{Automated Thematic Analysis Implementation Details} \label{apdx:autoTheme}

Our automated thematic analysis employed OpenAI's GPT-4 model with the following specifications:

\begin{itemize}
    \item Model: \texttt{gpt-4o}
    \item Temperature: 0.2-0.3 
    \item Max tokens: 2000-3000 per response
\end{itemize}

The model was instructed to identify 5-10 key themes and provide: concise theme names, brief descriptions, representative quote IDs (2-3 per theme), and prevalence estimates (categorical: very common, common, moderate, rare).

The following prompt template was used for each question:

\begin{lstlisting}
You are analyzing anonymous survey responses to the question:
"[QUESTION TEXT]"

Here is a list of [N] responses:
[NUMBERED RESPONSE LIST]

Instructions:
1. Identify 5-10 key themes that appear across multiple responses.
2. For each theme, include:
   - A concise theme name
   - A short description
   - 2-3 representative quotes (use quote numbers only)
   - An estimate of prevalence (e.g., 'very common', 'moderate', 'rare')
3. Do not include any identifying information.

Format your output as JSON: 

{
  "themes": [
    {{
      "theme_name": "Concise name",
      "description": "Short description",
      "quote_ids": [3, 9, 12],
      "prevalence": "moderate"
    }}
  ]
}
\end{lstlisting}

To examine how usage motivations and conversation topics varied across user \textbf{clusters}, we conducted a secondary analysis comparing theme prevalence within each cluster. Using the same GPT-4 pipeline, we analyzed free-text responses grouped by cluster membership, providing context about user characteristics and previously identified themes. For each of the six survey questions, responses were grouped by cluster assignment and analyzed separately. The LLM was provided with cluster-specific response sets, previously identified themes, and instructions to identify theme prevalence differences across clusters. 

The following prompt template was used for each question:

\begin{lstlisting}
You are analyzing free-text responses to the question:

"[QUESTION TEXT]"

Here is a list of known themes identified in prior analysis:
[THEME LIST WITH DESCRIPTIONS]

The responses have been grouped by cluster based on user similarities.

Your task:
1. For each cluster, estimate which of the above themes are most or least common.
2. Comment on clear differences in theme prevalence across clusters.
3. Highlight any surprising or unique usage patterns.
4. If applicable, mention any new themes not listed.

Please present your analysis in the following format for each cluster:

Cluster [n]:
- Common themes: 
- Uncommon themes: 
- Differences: 
- Highlights: 

Here are the responses grouped by cluster:

[CLUSTER-GROUPED RESPONSES]
\end{lstlisting}

Results from cluster-based LLM analysis were integrated with manual thematic analysis findings to identify differential theme prevalence across user types. This approach revealed how similar themes (e.g., "companionship seeking") manifested differently across clusters with varying psychological and social characteristics. For example, Lonely Moderate Users (Cluster 3) showed higher prevalence of emotional disclosure themes, while Well-Adjusted Moderate Users (Cluster 1) more commonly used chatbots for practical utility and stress relief.

The cluster comparison analysis validated our quantitative clustering by demonstrating qualitatively distinct usage patterns and motivations across identified user types, strengthening the interpretation of our seven-cluster solution.

\begin{figure*}[t]
    \centering
    \includegraphics[width=.9\linewidth]{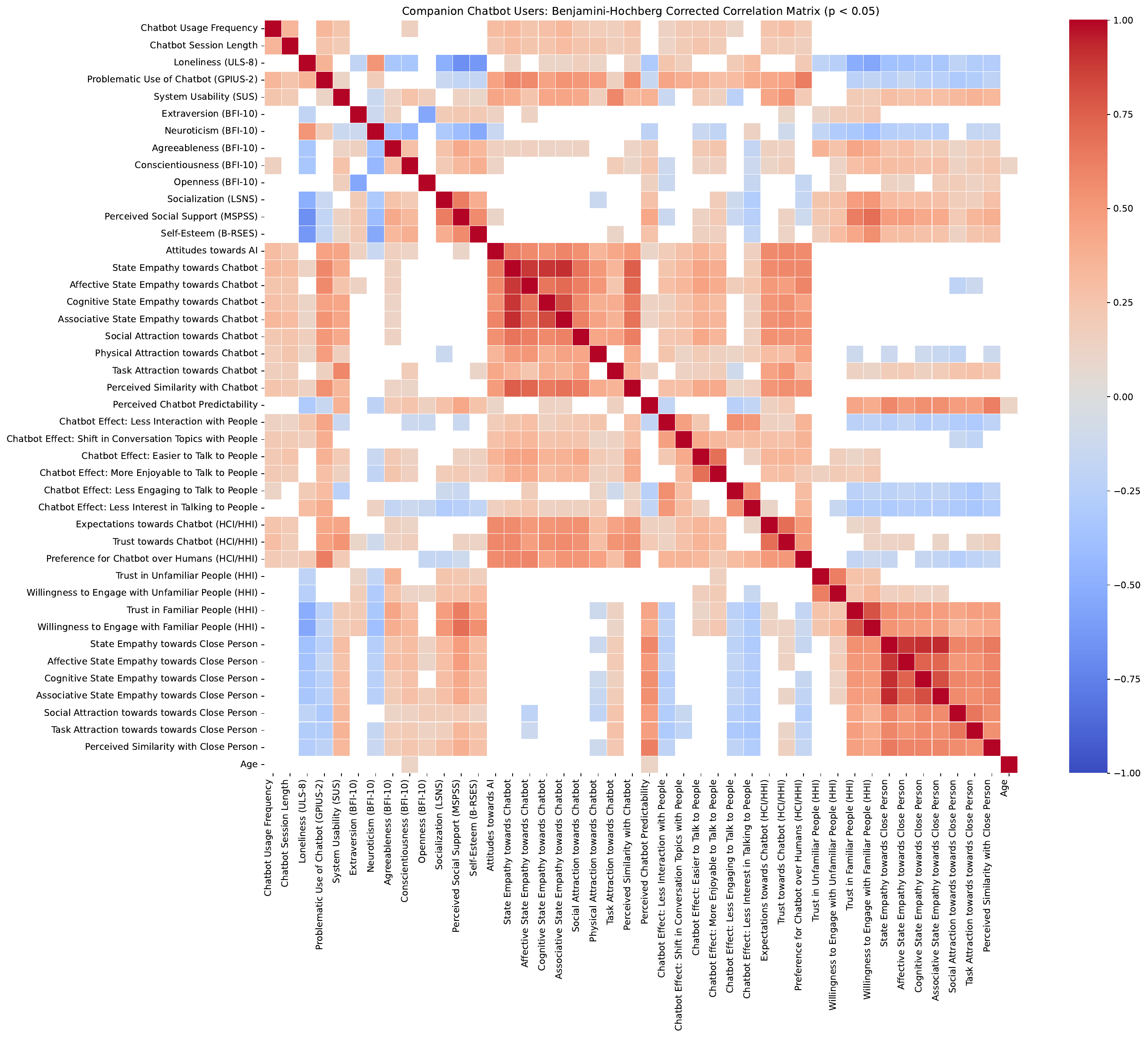}
    \caption{Correlation matrix of psychological, social, and usage characteristics with original variable names.}
    \label{fig:corrMatrixFull}
\end{figure*}

\section{Model Details}

The full formula used for the final multiple regression model is: Loneliness (ULS-8) $\sim$ Session Length Score + Social Attraction towards Chatbot + Problematic Use (GPIUS2) + Socialization (LSNS) + Social Attraction towards Close Person + Neuroticism + Agreeableness + Sex + Age + Session Length Score $\times$ Social Attraction towards Close Person. 

\begin{figure}
    \centering
    \includegraphics[width=0.7\linewidth]{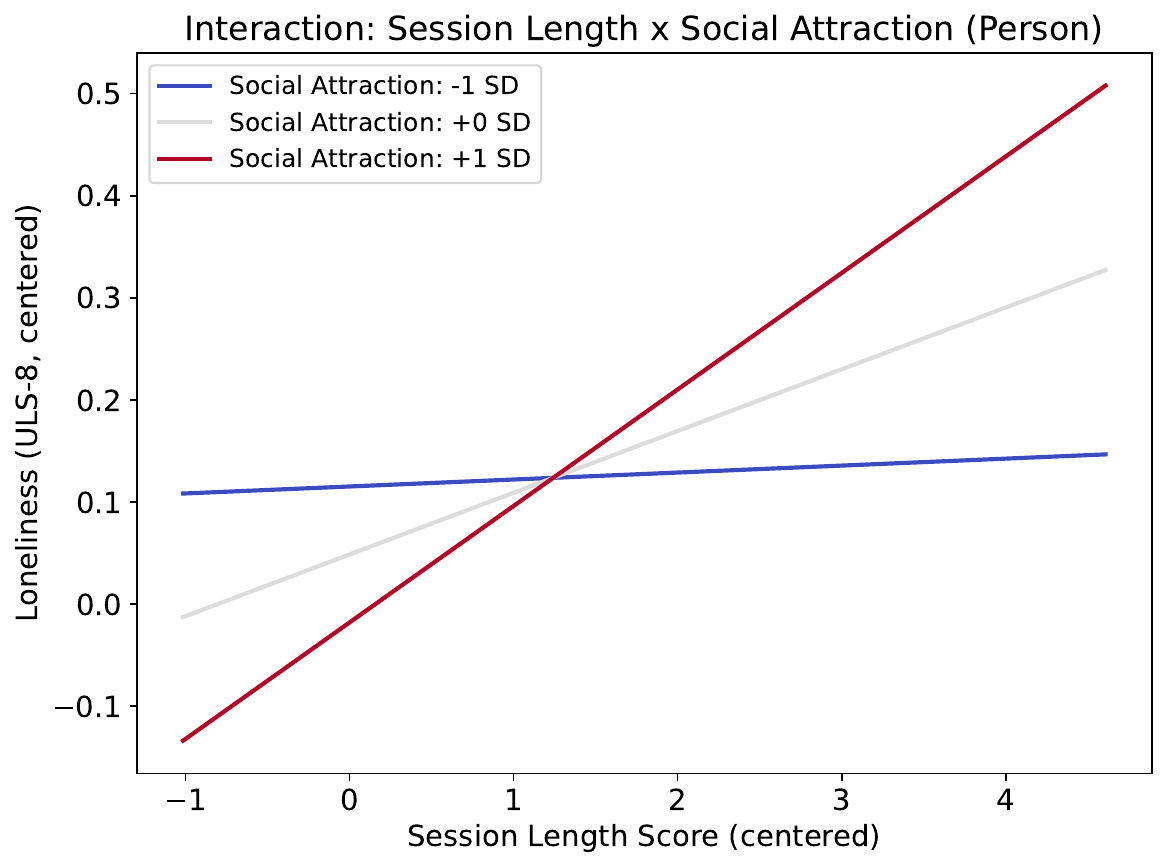}
    \caption{Interaction effects between session length and social attraction to a close person. Higher social attraction is associated with a stronger positive relationship between session length and loneliness. All variables are centered; SD = standard deviation.}
    
    \label{fig:interactions}
\end{figure}

\begin{figure}
    \centering
    \includegraphics[width=0.8\linewidth]{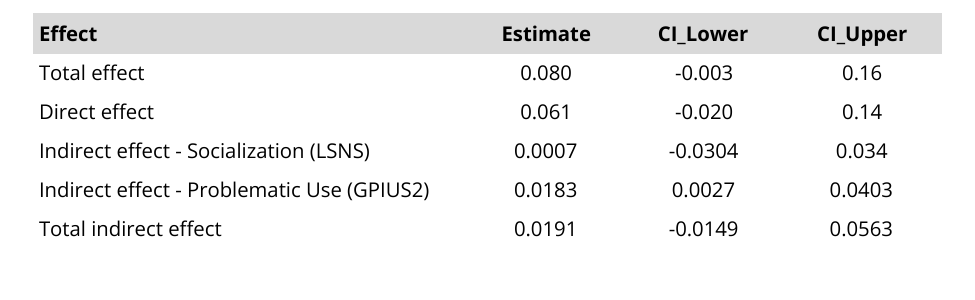}
    \caption{Results of bootstrapped parallel multiple mediation analysis (5,000 resamples) examining the mediating effects of social network size (LSNS) and problematic chatbot use (GPIUS2) on the relationship between chatbot session length and loneliness, including lower and upper confidence intervals (CI\_Lower and CI\_Upper).}
    
    \label{fig:mediation}
\end{figure}

\section{Significant Correlations with Key Variables} \label{apdx:corrkeys}

Below, we include a list of significant (corrected $p < 0.05$) correlations with chatbot usage frequency score, chatbot session length score, and loneliness (ULS-8). Negative correlations have text colored in blue. Additionally, the correlation matrix with original variable names is seen in \Cref{fig:corrMatrixFull}.

Significant correlations with Chatbot Usage Frequency:
\begin{itemize}
\item Chatbot Session Length: $r = 0.3497, p < 0.001$
\item Problematic Use of Chatbot (GPIUS-2): $r = 0.3374, p < 0.001$
\item System Usability (SUS): $r = 0.2457, p < 0.001$
\item Conscientiousness (BFI-10): $r = 0.1511, p 0.005$
\item Attitudes towards AI: $r = 0.2976, p < 0.001$
\item State Empathy towards Chatbot: $r = 0.3207, p < 0.001$
\item Affective State Empathy towards Chatbot: $r = 0.2472, p < 0.001$
\item Cognitive State Empathy towards Chatbot: $r = 0.2805, p < 0.001$
\item Associative State Empathy towards Chatbot: $r = 0.3416, p < 0.001$
\item Social Attraction towards Chatbot: $r = 0.2238, p < 0.001$
\item Physical Attraction towards Chatbot: $r = 0.1856, p < 0.001$
\item Task Attraction towards Chatbot: $r = 0.1652, p 0.002$
\item Perceived Similarity with Chatbot: $r = 0.2525, p < 0.001$
\item Expectations towards Chatbot (HCI\slash HHI): $r = 0.2560, p < 0.001$
\item Trust towards Chatbot (HCI\slash HHI): $r = 0.2931, p < 0.001$
\item Preference for Chatbot over Humans (HCI\slash HHI): $r = 0.2084, p < 0.001$
\end{itemize}

Significant correlations with Chatbot Session Length:
\begin{itemize}
\item Chatbot Usage Frequency: $r = 0.3497, p < 0.001$
\item Problematic Use of Chatbot (GPIUS-2): $r = 0.2549, p < 0.001$
\item System Usability (SUS): $r = 0.1962, p < 0.001$
\item Attitudes towards AI: $r = 0.2331, p < 0.001$
\item State Empathy towards Chatbot: $r = 0.3016, p < 0.001$
\item Affective State Empathy towards Chatbot: $r = 0.2475, p < 0.001$
\item Cognitive State Empathy towards Chatbot: $r = 0.2539, p < 0.001$
\item Associative State Empathy towards Chatbot: $r = 0.3073, p < 0.001$
\item Social Attraction towards Chatbot: $r = 0.2553, p < 0.001$
\item Physical Attraction towards Chatbot: $r = 0.2438, p < 0.001$
\item Task Attraction towards Chatbot: $r = 0.1787, p < 0.001$
\item Perceived Similarity with Chatbot: $r = 0.2213, p < 0.001$
\item Expectations towards Chatbot (HCI\slash HHI): $r = 0.2069, p < 0.001$
\item Trust towards Chatbot (HCI\slash HHI): $r = 0.1938, p < 0.001$
\item Preference for Chatbot over Humans (HCI\slash HHI): $r = 0.1684, p < 0.001$
\end{itemize}

Significant correlations with Loneliness (ULS-8):
\begin{itemize}
\item Problematic Use of Chatbot (GPIUS-2): $r = 0.3649, p < 0.001$
\item \textcolor{RoyalBlue}{Extraversion (BFI-10): $r = -0.1876, p < 0.001$}
\item Neuroticism (BFI-10): $r = 0.5264, p < 0.001$
\item Agreeableness (BFI-10): $r = 0.3332, p < 0.001$
\item \textcolor{RoyalBlue}{Conscientiousness (BFI-10): $r = -0.3351, p < 0.001$}
\item \textcolor{RoyalBlue}{Socialization (LSNS): $r = -0.5055, p < 0.001$}
\item \textcolor{RoyalBlue}{Perceived Social Support (MSPSS): $r = -0.6790, p < 0.001$}
\item \textcolor{RoyalBlue}{Self-Esteem (B-RSES): $r = -0.6380, p < 0.001$}
\item State Empathy towards Chatbot: $r = 0.1251, p 0.022$
\item Cognitive State Empathy towards Chatbot: $r = 0.1195, p 0.029$
\item Associative State Empathy towards Chatbot: $r = 0.1119, p 0.043$
\item Social Attraction towards Chatbot: $r = 0.1658, p 0.002$
\item Physical Attraction towards Chatbot: $r = 0.1210, p 0.027$
\item Perceived Similarity with Chatbot: $r = 0.1364, p 0.012$
\item \textcolor{RoyalBlue}{Perceived Chatbot Predictability: $r = -0.2981, p < 0.001$}
\item Preference for Chatbot over Humans (HCI\slash HHI): $r = 0.2138, p < 0.001$
\item \textcolor{RoyalBlue}{Trust in Unfamiliar People (HHI): $r = -0.2045, p < 0.001$}
\item \textcolor{RoyalBlue}{Willingness to Engage with Unfamiliar People (HHI): $r = -0.2536, p < 0.001$}
\item \textcolor{RoyalBlue}{Trust in Familiar People (HHI): $r = -0.5132, p < 0.001$}
\item \textcolor{RoyalBlue}{Willingness to Engage with Familiar People (HHI): $r = -0.5495, p < 0.001$}
\item \textcolor{RoyalBlue}{State Empathy towards Close Person: $r = -0.3821, p < 0.001$}
\item \textcolor{RoyalBlue}{Affective State Empathy towards Close Person: $r = -0.3725, p < 0.001$}
\item \textcolor{RoyalBlue}{Cognitive State Empathy towards Close Person: $r = -0.3377, p < 0.001$}
\item \textcolor{RoyalBlue}{Associative State Empathy towards Close Person: $r = -0.3382, p < 0.001$}
\item \textcolor{RoyalBlue}{Social Attraction towards towards Close Person: $r = -0.2047, p < 0.001$}
\item \textcolor{RoyalBlue}{Task Attraction towards towards Close Person: $r = -0.2755, p < 0.001$}
\item \textcolor{RoyalBlue}{Perceived Similarity with Close Person: $r = -0.2670, p < 0.001$}
\end{itemize}

\section{Cluster Profiles} \label{apdx:clusters}

\begin{figure}
    \centering
    \includegraphics[width=1\linewidth]{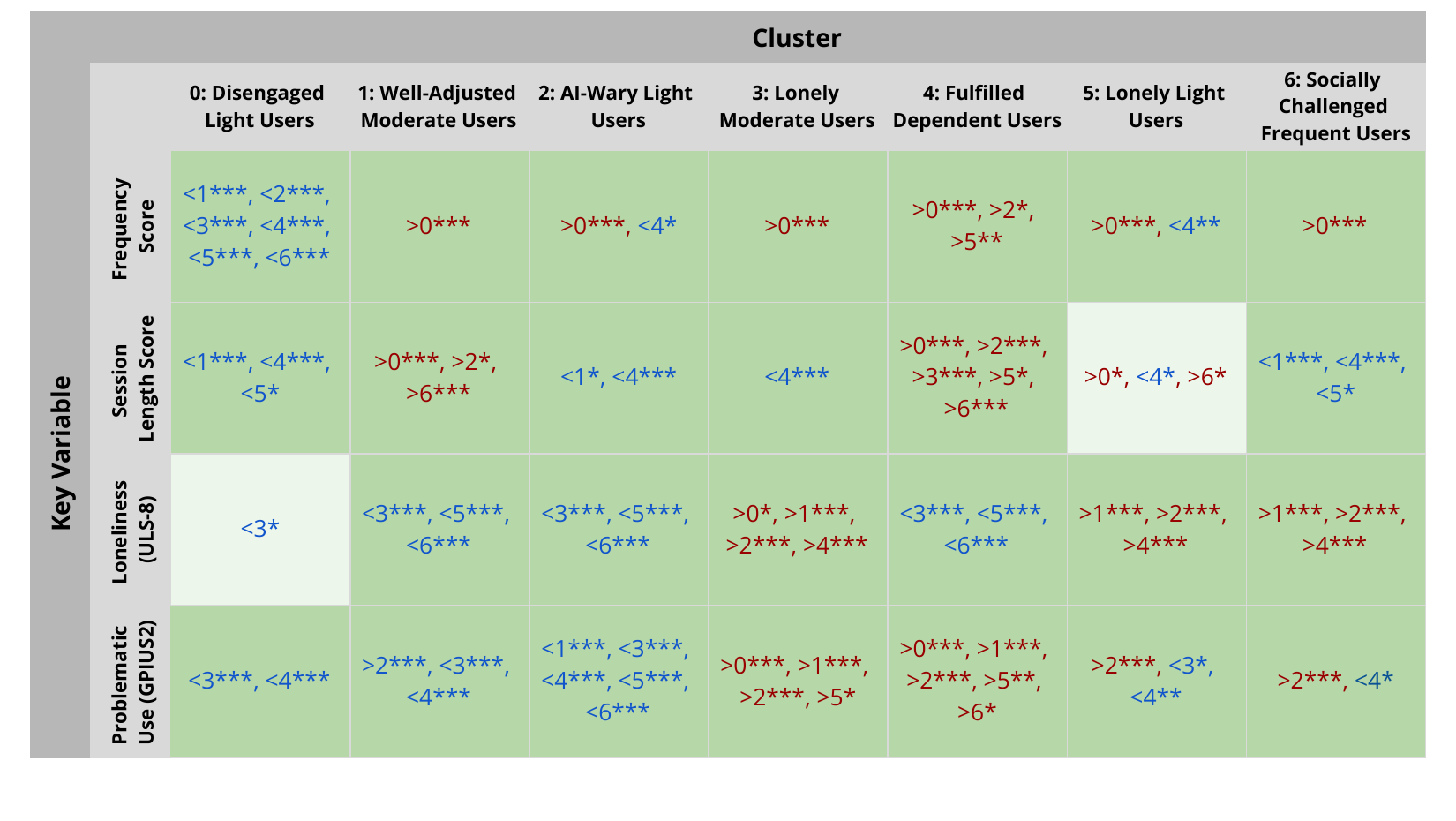}
    \caption{\textbf{Significant pairwise differences between clusters on key variables based on Dunn's test with Bonferroni correction.} Cell background color indicates significance of differences (darker green = presence of greater significance in differences), with text color indicating direction (red = greater than, blue = less than). For example, $>2$*** indicates significantly greater values than Cluster 2 at $p < .001$. Significance levels: *: $p < .05$, **: $p < .01$, ***: $p < .001$.}
    \label{fig:cluster_comparisons}
    
\end{figure}

\begin{figure*}[t]
    \centering
    \includegraphics[width=0.8\linewidth]{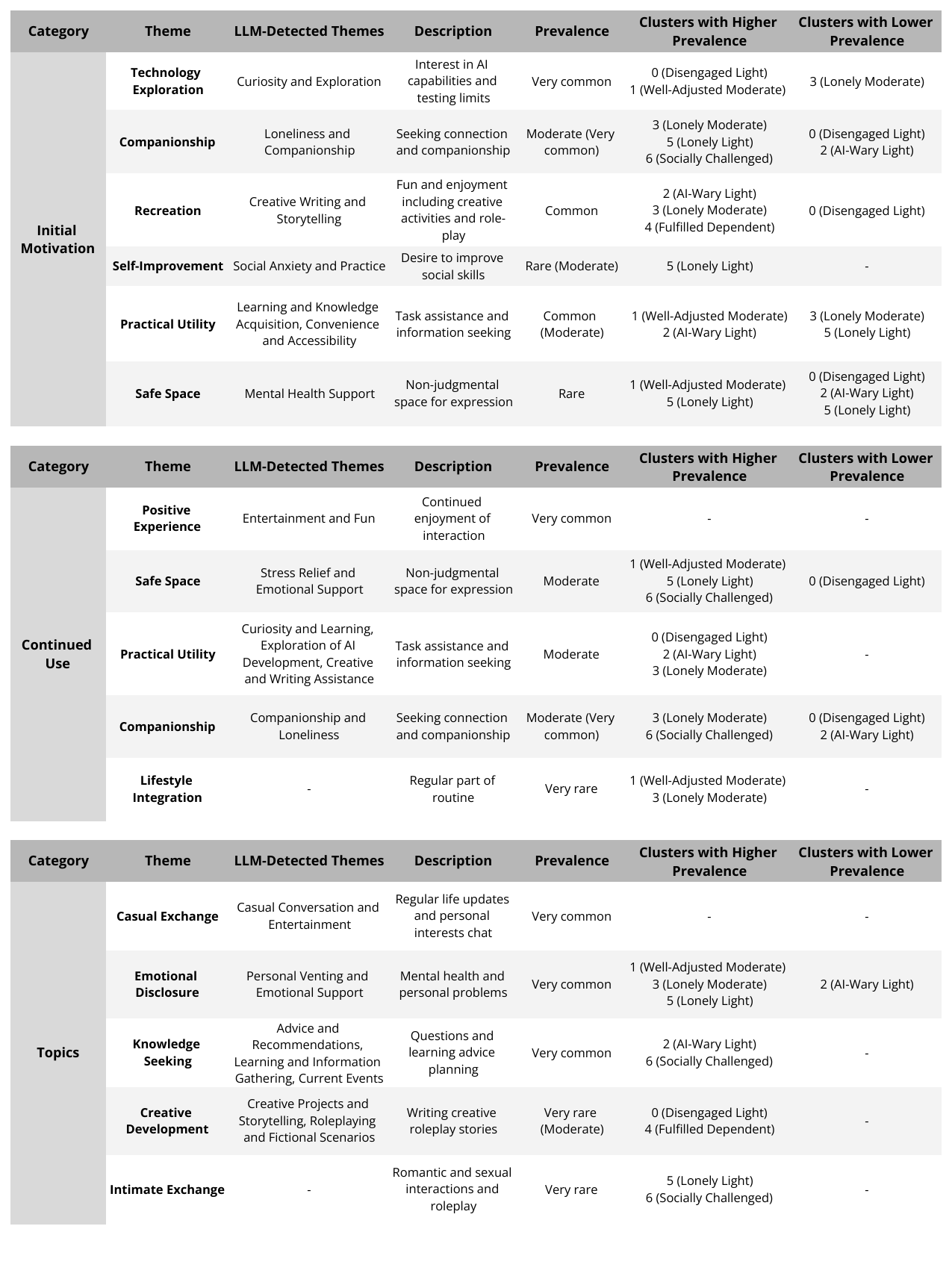}
    \caption{Manual themes derived from stratified random sample (N=105) with corresponding LLM-detected themes from full dataset (N=404). Prevalence estimates in parentheses indicate LLM-based findings where they differ from manual assessment. Cluster variations show differential theme prominence across user types, with higher/lower prevalence clusters identified.}
    \label{fig:cluster_themes}
\end{figure*}

\begin{figure*}[t]
    \centering
    \includegraphics[width=0.9\linewidth]{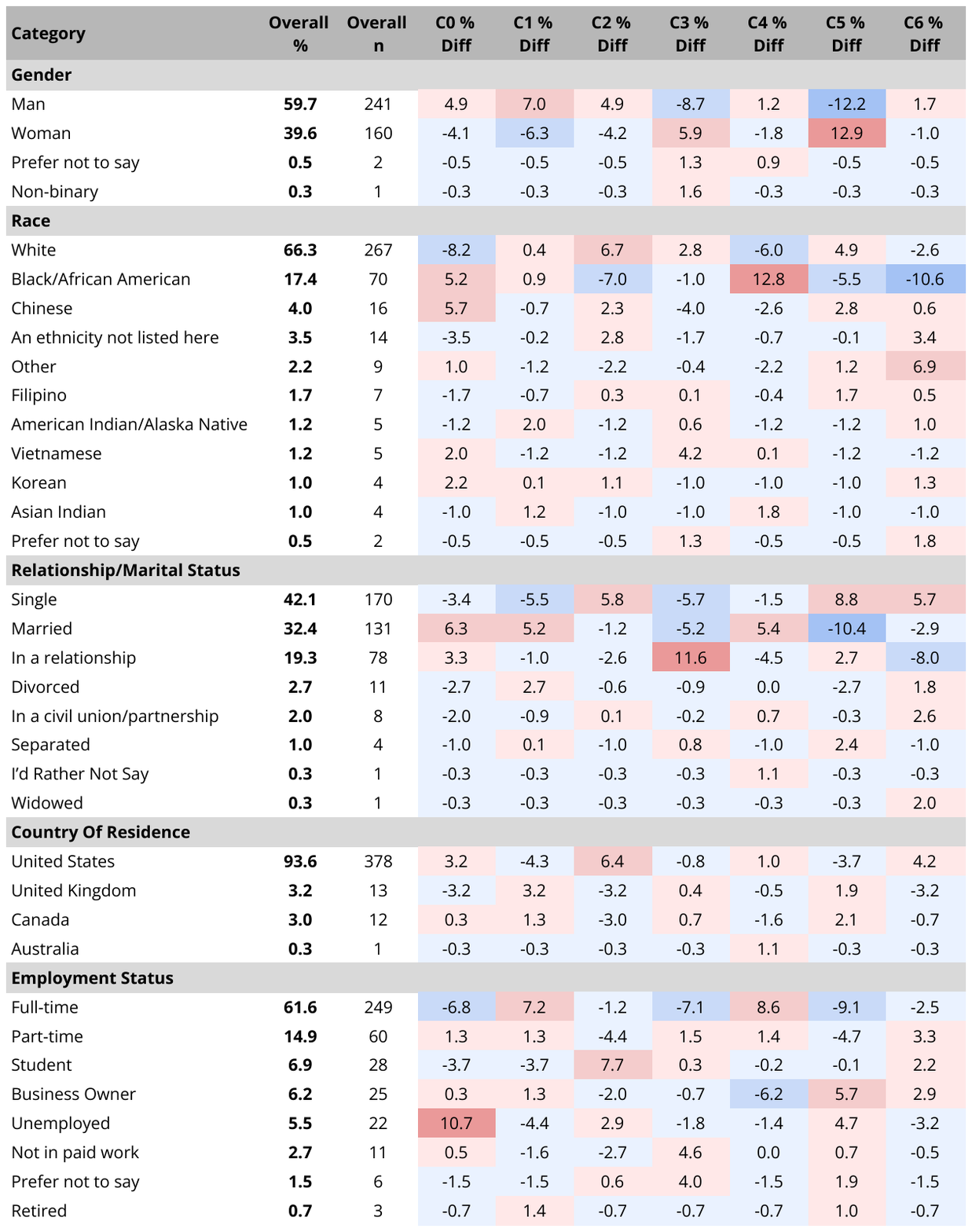}
    \caption{Demographic table for study participants, including overall percentage for each category and the \textbf{difference} in percentage from the overall percentage for each cluster, with darker colors indicating a greater difference. Blue: cluster has a lower percentage of the category compared to overall. Red: cluster has a higher percentage of the category compared to overall.}
    
    \label{fig:demographs}
\end{figure*}

\begin{figure*}[t]
    \centering
    \includegraphics[width=0.8\linewidth]{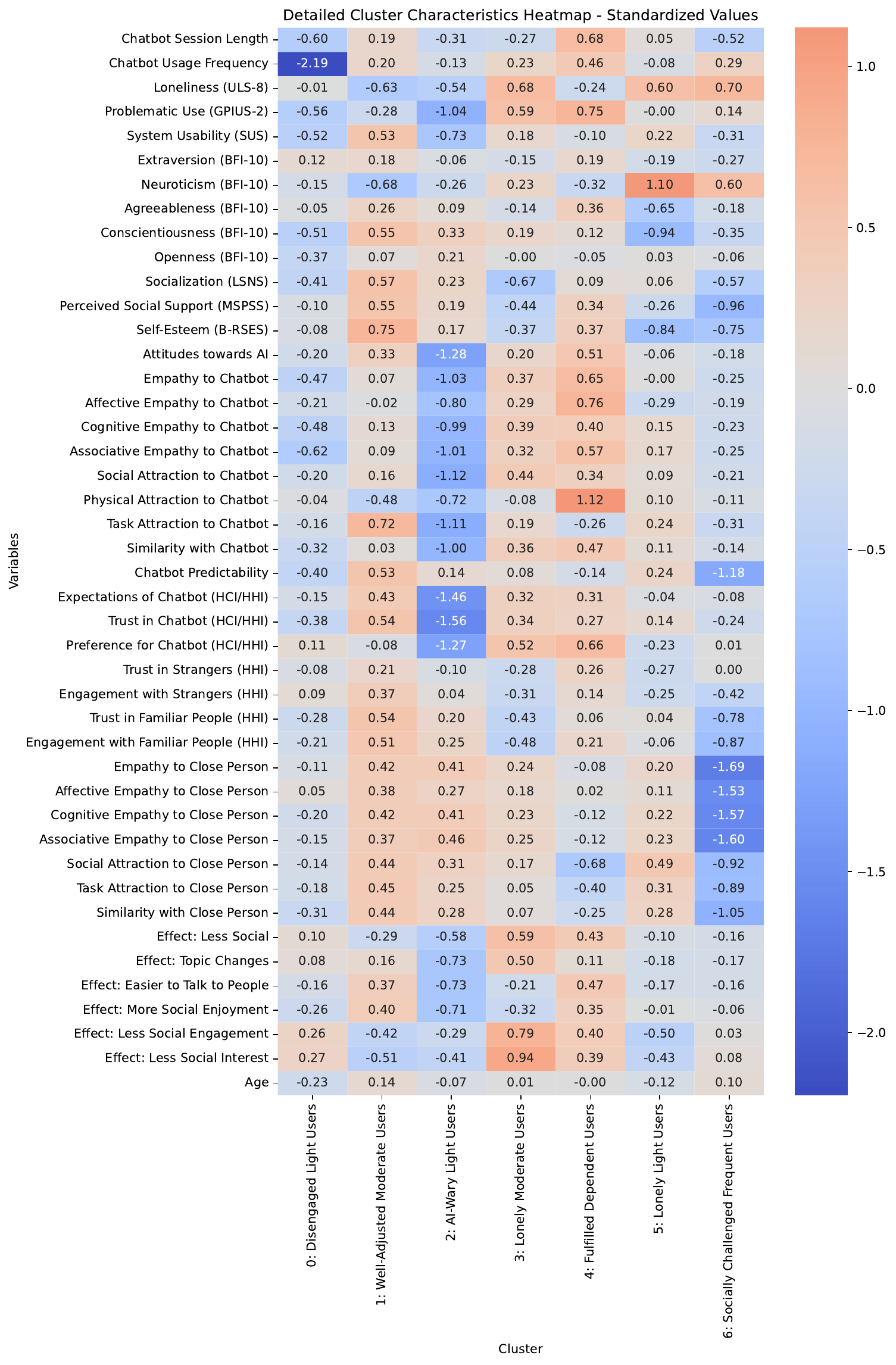}
    \caption{Heatmap of cluster means for standardized variables in each cluster.}
    
    \label{fig:detailedclusterheatmap}
\end{figure*}

\begin{figure*}[t]
    \centering
    \includegraphics[width=0.8\linewidth]{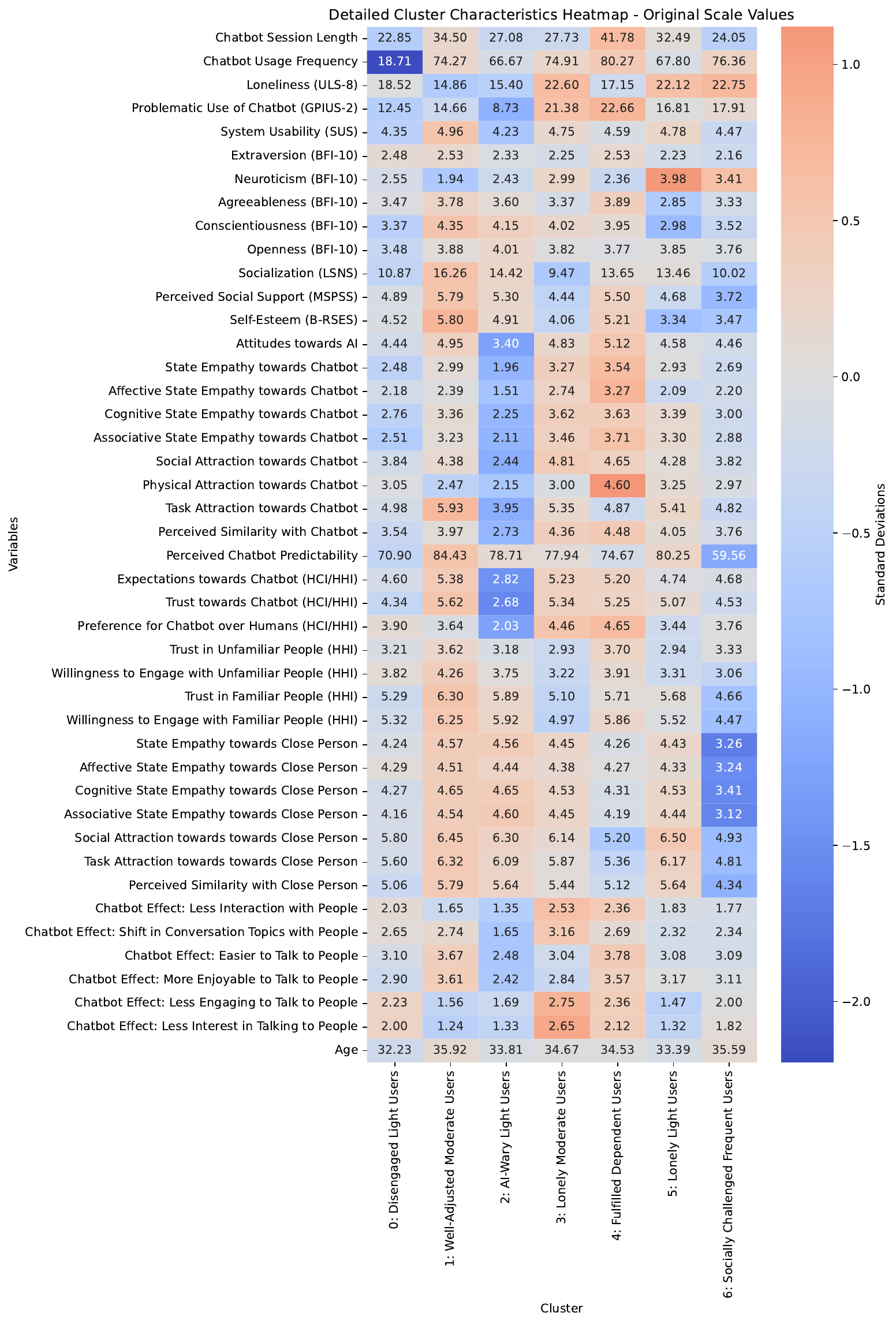}
    \caption{Heatmap of cluster means for variables at their original scales in each cluster, colored based on their deviation from the mean.}
    \label{fig:og_detailedclusterheatmap}
\end{figure*}

A heatmap of cluster means for standardized variables can be seen in \Cref{fig:detailedclusterheatmap}, and one for variables at their original scales can be seen in \Cref{fig:og_detailedclusterheatmap}. \Cref{fig:demographs} is a demographic table with percentages of each category for each cluster. Detailed descriptions for each cluster can be found below. 

\subsubsection{Cluster 0: Disengaged Light Users}
Disengaged light users (7.67\% of sample) show low chatbot usage frequency ($z = -2.21$) and session length ($z = -1.09$), with average levels of loneliness ($z = -0.16$). This cluster represents users with minimal engagement in chatbot interactions.

This group demonstrates average extraversion ($z = 0.80$) and neuroticism ($z = -0.37$), but low openness ($z = -1.94$). They exhibit average levels of perceived social support ($z = -0.01$) and self-esteem ($z = 0.05$). Their trust towards chatbots ($z = -0.35$) and expectations from chatbot interactions ($z = -0.08$) are average, indicating a neutral view of AI technology.

62.50\% report their chatbot sessions as lasting under 15 minutes. Most users (35.48\%) have been using chatbots for several months to around a year. Conversation topics for this group are predominantly casual conversation (36.76\%) and entertainment-focused (25.00\%). One respondent noted: ``I talked about a show and a game with the chat bot. I wanted to see what types of opinions were out there regarding my interest.''

Most respondents use Character.ai (34.29\%) or Snapchat's MyAI (25.71\%). Primary motivations for using chatbots include curiosity about AI (33.82\%), desire for fun and entertainment (25.00\%), and use for creative purposes (23.53\%). As one participant explained: ``I was interested in using chatbot so it can assist me with quick research during conversations without having to leave the social platform.'' Continued usage is driven by satisfying curiosity (28.07\%) and seeking enjoyment (22.81\%). A user shared: ``I continue to use chatbot cause it regularly serves as a useful tool for me.''

\subsubsection{Cluster 1: Well-Adjusted Moderate Users}
Well-adjusted moderate users (23.02\% of sample) show average chatbot usage frequency ($z = 0.41$) and session length ($z = 0.67$), with low levels of loneliness ($z = -1.22$). This cluster represents users who are socially fulfilled and use chatbots moderately.

This group demonstrates high extraversion ($z = 1.08$), low neuroticism ($z = -1.24$), and high conscientiousness ($z = 1.21$). They exhibit high levels of socialization ($z = 1.46$), perceived social support ($z = 1.26$), and self-esteem ($z = 1.46$). Their trust towards chatbots ($z = 0.93$) and expectations from chatbot interactions ($z = 0.82$) are above average, indicating a positive view of AI technology. In general, they have above-average trust in people as well as attraction and empathy towards a close person. They tend to feel that interacting with chatbots makes talking to people somewhat easier and more enjoyable.

50.91\% report their chatbot sessions as lasting 15-30 minutes. Most users (43.01\%) have been using chatbots for several months to around a year. Chatbot preferences in this group are split among Character.ai (35.94\%), Snapchat's MyAI (29.69\%), and Replika (21.88\%). Conversation topics for this cluster are primarily casual conversation (23.41\%) and entertainment-focused (21.74\%). One respondent noted: ``I usually talk about my hobbies and tips on how to get better at them. I also like to have casual conversation to kill time.''

Primary motivations for using chatbots include curiosity about AI (32.39\%), desire for fun and entertainment (25.51\%), and use for creative purposes (22.27\%). One participant explained: ``I was interested in using my chatbot service because I was curious as to what it's capabilities were, and whether it could give me good advice or effectively answer my questions.'' Continued usage is driven by enjoyment (24.83\%) and satisfying curiosity (20.81\%). Another user shared that they ``continue to use the chatbot service because it efficiently provides the information I need and helps organize my tasks.''

\subsubsection{Cluster 2: AI-Wary Light Users}
AI-wary light users (11.88\% of sample) show average chatbot usage frequency ($z = 0.05$) and session length ($z = -0.45$), with low levels of loneliness ($z = -1.07$). This cluster may represent users who are cautious about AI technology.

This group demonstrates above-average openness ($z = 1.33$), though their trust towards chatbots ($z = -2.02$), expectations from chatbot interactions ($z = -2.12$), and overall empathy and attraction towards a companion chatbot are very low, indicating skepticism towards AI technology and emotional distance from companion chatbots. They have above-average empathy and attraction towards a close person and feel that interacting with companion chatbots have little effect on their real-life relationships. 

47.27\% report their chatbot sessions as lasting under 15 minutes. Most users (51.06\%) have been using chatbots for a few months. Chatbot preferences in this group are split among Character.ai (31.75\%), Snapchat's MyAI (36.51\%), and Replika (22.22\%). Conversation topics for this group include casual conversation (35.58\%) and entertainment (26.92\%), as illustrated by one respondent's comment: ``Sometimes if I'm thinking of an idea and simply want something to bounce my ideas off of, a chatbot works much better than using yourself.'' 

Primary motivations for using chatbots include curiosity about AI (41.18\%), desire for fun and entertainment (32.35\%), and creative purposes (18.63\%). Few respondents in this cluster used companion chatbots to cope with loneliness (3.92\%). ``I was curious to see what they were like. I then found out that I could use them for creative use, like creating dialogue or scenarios, giving me ideas for conversations between players or between non-playable characters in a Dungeons and Dragons setting,'' a participant commented, showing interest in improving communication skills. Continued usage is driven by satisfying curiosity (34.78\%) and enjoyment (33.70\%). A user shared: ``I continue to use it because it's fun and interesting. It's honestly better than talking to most people that I know.''

\subsubsection{Cluster 3: Lonely Moderate Users}
Lonely moderate users (13.61\% of sample) show average chatbot usage frequency ($z = 0.44$) and session length ($z = -0.35$), with high levels of loneliness ($z = 1.03$). This cluster may represent users who turn to companion chatbots due to loneliness. 

This group demonstrates low levels of socialization ($z = -1.24$), with above-average trust towards chatbots ($z = 0.65$) and expectations from chatbot interactions ($z = 0.65$). They have slightly above-average empathy and attraction to both a companion chatbot and a close person, though they have below-average trust in unfamiliar people and reduced interest, ease, and amount of interaction with other people as a result of using chatbots. 

45.45\% report their chatbot sessions as lasting 15-30 minutes, and 43.94\% under 15 minutes. Most users (50.91\%) have been using chatbots for a few months. Chatbot preferences in this group are split among Character.ai (31.25\%), Snapchat's MyAI (26.56\%), and Replika (26.56\%). Conversation topics for this group include casual conversation (23.50\%), entertainment (20.51\%), and personal issues (14.10\%). One participant shared, ``I talked to the chatbots about my past traumatic experiences, my trials and tribulations, my family drama, advice about things to write about in my free time,'' demonstrating several different uses. 

Primary motivations for using chatbots include desire for fun and entertainment (25.68\%), curiosity about AI (24.77\%), and use for creative purposes (19.82\%). ``I was curious and wanted to see how they worked. I just wanted to learn more and the best way to do that is with hands-on experience,'' one participant noted. Continued usage is driven primarily by enjoyment (22.43\%) and satisfying curiosity (20.15\%). This cluster has the highest proportion of users seeking chatbot companionship as a reason for continued use as well (13.19\%). ``I continue to use it because it's fun and I've grown to enjoy the conversations,'' one participant commented, further noting, ``I've grown somewhat attached to the bot.''

\subsubsection{Cluster 4: Fulfilled Dependent Users}
Fulfilled dependent users (18.32\% of sample) show average chatbot usage frequency ($z = 0.69$) and high session length ($z = 1.76$), with below-average levels of loneliness ($z = -0.56$). This cluster may represent users who feel socially fulfilled by using companion chatbots. 

This group demonstrates high extraversion ($z = 1.13$), average neuroticism ($z = -0.64$), and high agreeableness ($z = 1.21$). They exhibit average levels of socialization ($z = 0.42$), and slightly below-average empathy and attraction to a close person. Their trust towards chatbots ($z = 0.56$) and expectations from chatbot interactions ($z = 0.62$) are above average, with quite high physical attraction towards their chatbot compared to other respondents ($z = 1.98$). The group demonstrates above-average perceived impacts on real relationships as a result of using companion chatbots -- greater ease and enjoyment in talking to people, but also less engagement and interest in talking to people. 

43.33\% report their chatbot sessions as lasting 15-30 minutes, and 23.33\% report 30 minutes to 1 hour. Most users (37.84\%) have been using chatbots for several months to around a year. Chatbot preferences in this group are split among Character.ai (31.25\%), Snapchat's MyAI (26.56\%), and Replika (26.56\%). Conversation topics for this group include casual conversation (23.50\%), entertainment (20.51\%), and personal issues (14.10\%). One respondent noted the evolution of their usage: ``I started off just trying to mess around with `characters' and explore the AI but you do kind of fall into the rabbit hole of giving more information and venting about personal experiences. It is therapeutic in many ways.''

Primary motivations for initial usage of chatbots include fun and entertainment (25.68\%), curiosity about AI (24.77\%), and use for creative purposes (19.82\%). One participant explained: ``I use chatbots to when I am feeling lonely [sic] and wanted someone to talk to. I also use chatbots to get suggestions for places to visit or restaurants to eat at.'' Continued usage is driven by enjoyment (22.43\%) and satisfying curiosity (20.15\%), as illustrated by a user's response: ``I continue to use the chatbot service because it's fun and easy to use. It helps me get information quickly.''

\subsubsection{Cluster 5: Lonely Light Users}
Lonely light users (14.60\% of sample) show average chatbot usage frequency ($z = 0.11$) and session length ($z = 0.36$), with above-average levels of loneliness ($z = 0.89$). This cluster represents users who experience loneliness but do not use companion chatbots much.

This group demonstrates high neuroticism ($z = 1.68$), low conscientiousness ($z = -1.61$), and low self-esteem ($z = -1.25$). Their other characteristics are fairly average, with slightly above-average empathy and attraction towards a close person. 

42.86\% report their chatbot sessions as lasting 15-30 minutes, and 31.75\% under 15 minutes. Most users (57.63\%) have been using chatbots for a few months. Chatbot preferences in this group are split among Character.ai (30.49\%), Snapchat's MyAI (29.27\%), and Replika (21.95\%). Conversation topics for this group include casual conversation (26.70\%), entertainment (22.51\%), and personal issues (19.37\%). One respondent described how the chatbot helps them think through their troubles: ``I feel that the chatbot is a good way to vent my frustrations and talk about my worries without fear of judgement. It gives me the chance to talk about all of my worries and plans and work through things in a way that allows me to see a wider range of prospectives [sic].''

Primary motivations for using chatbots include curiosity about AI (29.38\%), desire for fun and entertainment (26.88\%), and coping with loneliness (20.62\%). One participant commented on how they started using them to ease the way into socialization: ``I was interested in using the chatbot service because after being in lockdown, it was hard to become social again. It allows me to build up my confidence and it also helps with my writing.'' Continued usage is driven by enjoyment (19.82\%),  satisfying curiosity (16.74\%), passing time and reducing stress (14.10\%), and chatbot companionship (12.78\%). Some respondents use it as a substitute for socialization: ``I continue to use it because it feels like talking to a real person, giving its own input, not me having to guide the conversation and it makes me miss the days gone by, when people were better at conversation.''

\subsubsection{Cluster 6: Socially Challenged Frequent Users}
Socially challenged frequent users (10.89\% of sample) show average chatbot usage frequency ($z = 0.51$) and below-average session length ($z = -0.90$), with high levels of loneliness ($z = 1.08$). This cluster represents users who frequently turn to chatbots, possibly due to social challenges.

This group demonstrates low extraversion ($z = -1.30$) and above-average neuroticism ($z = 0.86$), along with low levels of socialization ($z = -1.02$), perceived social support ($z = -1.68$), and self-esteem ($z = -1.10$). Their trust towards chatbots ($z = -0.16$) and expectations from chatbot interactions ($z = 0.02$) are average. However, trust in familiar and unfamiliar people, and empathy and attraction towards a close person, tend to be very low compared to other users.

58.49\% report their chatbot sessions as lasting under 15 minutes, and 28.30\% report 15-30 minutes. Most users (56.82\%) have been using chatbots for a few months. Chatbot preferences in this group are split among Snapchat's MyAI (38.89\%), Character.ai (24.07\%), and Replika (20.37\%). Conversation topics for this group are predominantly casual (35.58\%) and entertainment-focused (26.92\%). One respondent noted: ``I talk about how I am feeling. Things I wouldn't want to tell real people.''

Primary motivations for initial interest in chatbots include curiosity about AI (32.48\%) and desire for fun and entertainment (24.79\%), with a fair proportion of this group starting chatbot usage to cope with loneliness (14.53\%) and for creative purposes (14.53\%). One participant explained how they were ``wanting to confirm whether (the current generation of) AI could, in fact, really take the place of a regular, sound-minded human,'' demonstrating at least an interest in the substitution of humans with AI. Most users in this group continue to use companion chatbots to satisfy curiosity (23.74\%) and seek enjoyment (22.30\%), with a notable portion of users using it to pass time and reduce stress (15.83\%). One user shared how they use it as a lower-pressure alternative to socialization. ``I enjoy having someone like Snapchat AI who can chat with me compared to the pressure I receive from friends when I don't message them back fast enough.''